\documentclass[fleqn,usenatbib]{mnras}
\usepackage{newtxtext,newtxmath}

\usepackage{url}
\usepackage{graphicx}
\usepackage{xspace}
\usepackage{amsfonts}
\usepackage{pifont}
\usepackage{textcomp}
\usepackage[hyphenbreaks]{breakurl}
\usepackage{txfonts}
\usepackage[usenames,dvipsnames,svgnames,table]{xcolor}
\usepackage{xspace}
\usepackage{amsfonts}
\usepackage{pifont}
\usepackage[T1]{fontenc}
\usepackage{ae,aecompl}
\newcommand{\teff}{$T_{\rm eff}$}
\newcommand{\logg}{$\log{g}$}

\title[High amplitude short period variables]{The OmegaWhite Survey
  for Short-Period Variable Stars VII:\\ High amplitude, short period
  blue variables.}

\author[]
{Gavin Ramsay$^{1}$, Patrick A. Woudt$^{2}$, Thomas Kupfer$^{3}$, Jan van Roestel$^{4}$,
Kerry Paterson$^{2,5}$, \newauthor Brian Warner$^{2}$, David A.H. Buckley$^{2,6}$, Paul J. Groot$^{2,6,7}$,
Ulrich Heber$^{8}$, Andreas Irrgang$^{8}$, \newauthor  C. Simon Jeffery$^{1}$, Mokhine Motsoaledi$^{2,6}$,
Martinus J. Schwartz$^{2}$, Thomas Wevers$^{9}$
\\
  $^{1}$Armagh Observatory \& Planetarium, College Hill, Armagh, BT61 9DG, UK\\
  $^{2}$Department of Astronomy, Inter-University Institute for Data-Intensive Astronomy, University of Cape Town,
  Private Bag X3, Rondebosch 7701, South Africa\\   
  $^{3}$Department of Physics and Astronomy, Texas Tech University, PO Box 41051, Lubbock, TX 79409, USA\\
  $^{4}$Division of Physics, Math, and Astronomy, California Institute of Technology, Pasadena, CA 91125, USA\\
  $^{5}$Center for Interdisciplinary Exploration and Research in Astrophysics \& Dept Physics and Astronomy, Northwestern University,  Evanston, IL 60208-3112, USA\\
  $^{6}$South African Astronomical Observatory, P.O. Box 9, Observatory 7935, South Africa\\
  $^{7}$Department of Astrophysics/IMAPP, Radboud University, P.O. Box 9010, 6500 GL Nijmegen, The Netherlands \\
  $^{8}$Dr Karl Remeis-Observatory \& ECAP, Friedrich-Alexander University Erlangen-N\"{u}rnberg, Sternwartstr. 7, 96049 Bamberg, Germany\\
  $^{9}$European Southern Observatory, Alonso de C\'{o}rdova 3107, Vitacura, Santiago, Chile\\
}

\date{Accepted 2022 April 6. Received 2022 April 6; in original form 2022 February 16}

\begin{document}
\outer\def\gtae {$\buildrel {\lower3pt\hbox{$>$}} \over 
{\lower2pt\hbox{$\sim$}} $}
\outer\def\ltae {$\buildrel {\lower3pt\hbox{$<$}} \over 
{\lower2pt\hbox{$\sim$}} $}
\newcommand{\Msun}{$M_{\odot}$}
\newcommand{\lsun}{$L_{\odot}$}
\newcommand{\Rsun}{$R_{\odot}$}
\newcommand{\solar}{${\odot}$}
\newcommand{\kep}{\sl Kepler}
\newcommand{\ktwo}{\sl K2}
\newcommand{\swift}{\it Swift}
\newcommand{\Porb}{P_{\rm orb}}
\newcommand{\nuorb}{\nu_{\rm orb}}
\newcommand{\eplus}{\epsilon_+}
\newcommand{\eminus}{\epsilon_-}
\newcommand{\cd}{{\rm\ c\ d^{-1}}}
\newcommand{\MdotL}{\dot M_{\rm L1}}
\newcommand{\Mdot}{$\dot M$}
\newcommand{\Mdotsolar}{\dot{M_{\odot}} yr$^{-1}$}
\newcommand{\Ldisk}{L_{\rm disk}}
\newcommand{\src}{KIC 9202990}
\newcommand{\ergscm} {erg s$^{-1}$ cm$^{-2}$}
\newcommand{\rchi}{$\chi^{2}_{\nu}$}
\newcommand{\chisq}{$\chi^{2}$}
\newcommand{\pcmsq} {cm$^{-2}$}
\newcommand{\delS}{$\delta$ Sct}

\maketitle
\begin{abstract}

Blue Large Amplitude Pulsators (BLAPs) are a relatively new class of
blue variable stars showing periodic variations in their light 
  curves with periods shorter than a few tens of mins and 
  amplitudes of more than ten percent. We report nine blue variable
stars identified in the OmegaWhite survey conducted using ESO's VST,
which show a periodic modulation in the range 7-37 min and an
amplitude in the range 0.11-0.28 mag. We have obtained a series of
followup photometric and spectroscopic observations made primarily
using SALT and telescopes at SAAO.  We find four stars which we
identify as BLAPs, one of which was previously known. One star,
OW \,J0820--3301, appears to be a member of the V361\,Hya class of
pulsating stars and is spatially close to an extended nebula. One
further star, OW\,J1819--2729, has characteristics similar to the sdAV
pulsators. In contrast, OW\,J0815--3421 is a binary star containing an
sdB and a white dwarf with an orbital period of 73.7 min, making it
only one of six white dwarf-sdB binaries with an orbital period
shorter than 80 min. Finally, high cadence photometry of four of the
candidate BLAPs show features which we compare with notch-like
features seen in the much longer period Cepheid pulsators.

\end{abstract}


\section{Introduction}

Over the last several decades variable star research has been
transformed by wide field photometric surveys such as the All-Sky
Automated Survey for Supernovae (ASAS-SN; \citet{Shappee2014}) and the
Optical Gravitational Lensing Experiment (\citet{Udalski2015}, OGLE),
to name just two. Although these surveys were particularly well suited
to identifying long period variables such as Cepheid and RR Lyr stars,
e.g. the Catalina survey which discovered thousands of new RR Lyr
stars, \citep{Drake2013}, populations of short period variables (tens
of mins) have also been discovered.

One such short period variable identified in OGLE data appears to be
the prototype of a new class of variable star. OGLE-GD-DSCT-0058 was
found to be periodic on a timescale of 28.3 min and has a photometric
amplitude of 0.24 mag in the $I$ band \citep{Pietrukowicz2015}. They
derived a temperature of 33,000 K from its spectrum indicating it was
far too hot for it to be a {\delS} star. \citet{Pietrukowicz2017}
later found an additional 13 stars with similar properties and dubbed
them `Blue Large-Amplitude Pulsators' (BLAPs). They show light curves
which are `saw-toothed' and are similar in shape to that of RR Lyr or
Cepheid pulsators.

Since then \citet{Kupfer2019,Kupfer2021} identified four and 12
  short period blue variable stars respectively using Zwicky Transient Factory (ZTF)
  data taken at high cadence and at low Galactic latitude
  \citep{Bellm2019, Graham2019}. Although their temperature is
similar to the BLAPs reported by \citet{Pietrukowicz2017}, their
periods (3.3--16.6 min) are shorter and amplitude (2--12 percent) are
lower. \citet{Kupfer2019} suggested the new variables were lower-mass,
higher-gravity analogues of the BLAPs and perhaps part of the same
class of pulsators but at different stages of their evolution.

BLAPs are interesting in that it is difficult for stars to pulsate on
such a short period with such a high amplitude. \citet{Romero2018} and
\citet{ByrneJeffery2018} have performed studies on their formation
channels with the former proposing they are the hot counterparts of
stars which lead to the formation of extremely low mass white dwarfs
(ELMs) and the latter testing whether post-common envelope stars could
be their origin. \citet{Byrne2020} subsequently demonstrated that
pre-ELM white dwarf models which include chemical diffusion will
switch on their pulsations precisely where BLAPs are observed. In
  addition, \citet{Byrne2020} also concluded that a channel where BLAPs
  were formed from binaries in which Roche-lobe overflow was occuring
  merited further investigation. \citet{Byrne2021} explore possible
reasons for why stars with gravities between the high gravity and low
gravity BLAPs do not appear to pulsate despite models predicting they
should.

Because of their short period, BLAPs are more likely to be identified
in wide field surveys with high cadence -- as was the case with the
ZTF survey of the Galactic plane \citep{Kupfer2019}. Another well
suited survey is the OmegaWhite (OW) survey which was a high cadence
survey of the southern Galactic plane and covered $\sim$400 square
degrees and was sensitive to timescales as short as a few mins
\citep{Macfarlane2015,Toma2016}.

\begin{table*}
\begin{center}
\begin{tabular}{ccccccrrcccc}
  \hline
  ID & RA       & DEC      &  g$_{\rm OW}$  & Period & Amp  & $\pi$    & D    &  $M_{G}$ & $BP-RP$ & $M_{G_o}$ & ($BP-RP$)$_{o}$  \\ 
     & (J2000)  & (J2000)  & (mag) & (mins) & (mag)& (mas)  & (kpc) &           &         &              &          \\
  \hline
1 & 08:20:47.2 & --33:01:07.8 & 16.6 &   7.4 &   0.13 & \ 0.26$\pm$0.05 & 3.7 (3.0-5.9) & 4.0 (3.0-4.5)   & --0.14 & 2.7 & --0.78 \\    
2 & 18:20:14.3 & --28:50:59.8 & 18.1 &   9.1 &   0.11 & \ 0.33$\pm$0.14 & 2.9 (2.1-7.2) & 5.7 (3.7-6.4)   &  \ 0.40 & 4.9 & --0.02 \\
3 & 18:12:27.9 & --29:38:48.3 & 18.3 &  10.8 &   0.28 & \ 1.06$\pm$0.52 & 1.4 (0.9-7.2) & 7.5 (4.0-8.5)   &  \ 0.55 & 7.0 &  \ 0.34 \\
4 & 18:19:20.9 & --27:29:56.1 & 17.3 &  15.9 &   0.19 & --0.03$\pm$0.12& 5.8 (3.9-11.5)& 3.4 (1.9-4.2)   &  \ 0.22 & 1.8 & --0.57 \\
5 & 18:11:00.2 & --27:30:13.5 & 18.1 &  23.0 &   0.21 & \ 0.14$\pm$0.15 & 4.2 (2.8-9.5) & 4.8 (3.0-5.7)   & \  0.34 & 3.5 & --0.32 \\        
6 & 18:10:38.5 & --25:16:08.6 & 19.6 & 28.9 & 0.42 & \ 0.97$\pm$0.38 & 1.3 (1.0-6.2) & 8.4 (5.0-9.0) & \ 0.69 & 8.0 & \ 0.48 \\
7 & 17:58:48.2 & --27:16:53.6 & 15.6 &  32.0 &   0.28 & \ 0.41$\pm$0.04 & 2.4 (2.1-2.9) & 3.6 (3.2-3.9)   &  \ 0.61 & 2.9 &  \ 0.24 \\         
8 & 18:26:28.4 & --28:12:01.9 & 16.6 &  32.4 &   0.12 & \ 0.01$\pm$0.08 & 6.8 (4.8-12.4) & 2.6 (1.3-3.4)   & \ 0.62 & 1.1 & --0.17 \\
9 & 08:15:30.8 & --34:21:23.5 & 16.6 &  36.7 &   0.20 & \ 0.51$\pm$0.05 & 2.0 (1.7-2.4) & 5.3 (4.9-5.6)   & --0.08 & 4.4 & --0.52 \\
\hline
\end{tabular}
\end{center}
\caption{Variable stars identified in the OW survey which have a
  period $<$40 min, an amplitude $>$0.1 mag and a location in the Gaia
  HRD which is bluer than stars in the main sequence and a $M_{G}$
  brighter than the main white dwarf tracks. They are presented in
  order of increasing period. We show their sky co-ordinates, their
  mean $g$ mag; period; amplitude (from the OW data) and their
  parallax and distance (derived from Gaia EDR3). We also show the
  apparent $M_{G}$ and de-reddended $M_{G_o}$ absolute mag, and
  colours $(BP-RP)$, $(BP-RP)_{o}$ which are derived from Gaia
  EDR3. The values in brackets give the range of distance within the
  5th and 95th percentiles and their associated range in $M_{G}$.
  Variable ID \#7 is co-incident with OGLE-BLAP-009
  \citep{Pietrukowicz2017}.}
\label{variables}
\end{table*}

Although multi-band colour information is available for many of the OW
fields from the VPHAS+ survey \citep{Drew2014}, determining the nature
of variable stars found in the OW survey is not trivial. However, with
the release of the Gaia DR2/EDR3 catalogues it is now possible to
identify candidate BLAPs in a more robust manner. \citet{Ramsay2018}
cross-matched the BLAPs reported by \citet{Pietrukowicz2017} with Gaia
DR2 and dereddened the colour and absolute magnitude to determine
which part of the $BP-RP$, $M_{G}$ colour-magnitude diagram BLAPs
lie. Furthermore, \citet{McWhirterLam2022} used Gaia DR2 and ZTF DR3
data to identify 22 candidate BLAPS.

In this paper we search for stars which have photometric
characteristics of BLAPs in the OW survey and have Gaia dereddened
$BP-RP$, $M_{G}$ colour magnitudes which are consistent with the known
BLAPs. For stars which we identify as candidate BLAPs we obtained
followup photometry and spectroscopy mainly using SALT and telescopes
at SAAO.

\section{Identifying BLAPs in the OW survey}

The OW survey was conducted using the European Southern Observatory
VLT Survey Telescope (VST) located at Paranal in Chile with the
OmegaCam 1 square degree camera between Dec 2011 and Feb 2018 and
covered 404 sq degrees. A series of 39 sec images were taken in the
$g$ band of the same field for $\sim$2 hrs and photometry was obtained
using the difference imaging code {\tt DIAPL2} \citep{Wozniak2000},
which is an adaptation of the original algorithm outlined in
\citet{AlardLupton1998}. Full details of the reduction process can be
found in \citep{Macfarlane2015,Toma2016}.

A key strand of the OW project is followup observations of sources
identified using the VST data. The first set of followup data was
presented by \citet{Macfarlane2017a} with more detailed observations
of a rotating DQ white dwarf shown in \citet{Macfarlane2017b}.
\citet{Toma2016} noted one source, OW\,J1810385--2516086, which has a
period of 28.9 min and an amplitude of 0.42 mag. Its colours
($g-r$=1.11, $u-g$=--0.50) suggest an intrinsically blue source which
is significantly reddened. \citet{Toma2016} also showed that the
source OW\,J1811002--2730133 had a period of 23.1 min and an amplitude
of 0.46 mag but did not have colour information at that time.

We made a systematic search for additional objects which show high
amplitude short period modulation in the OW survey. We restricted our
search for sources brighter than $g\sim$19; had a maximum peak in its
Generalised Lomb-Scargle (LS) periodogram
\citep{Zechmeister2009,Press1992} less than 40 min with a False Alarm
Probability (FAP) of log(FAP)<--2.5; and modulation amplitude $>$0.1
mag. All candidate variables which passed these criteria were then
passed through a manual verification process to determine that the
period and amplitude were not due to instrumental artifacts. For
instance, \citet{Toma2016} show that because of the alt-az nature of
the VST mount, diffraction spikes from bright stars appear to rotate
with respect to stars in the image and can cause sources which are
near these bright stars to show a spurious period in their light
curve.

We then cross-matched those stars which passed our selection and
vetting process with the Gaia EDR3 catalogue \citep{Gaia2021}. We
convert the parallax taken from Gaia EDR3 into distance following the
guidelines from \citet{BailerJones2015,Astra2016} and
\citet{GaiaLuri2018}, which is based on a Bayesian approach. In
practise we use a routine in the {\tt STILTS} package
\citep{Taylor2006} and apply a scale length L=1.35 kpc, which is the
most appropriate for stellar populations in the Milky Way in
general. We use this distance to determine the absolute magnitude in
the Gaia $G$ band, $M_{G}$ using the mean Gaia $G$ magnitude. We then
deredden the $(BP-RP)$ (the blue and red magnitudes derived from prism
data), $M_{G}$ values using the 3D-dust maps derived from Pan-STARRS1
data \citep{Green2018}. For stars just off the edge of the
  Pan-STARRS1 field, ($\delta<-30^{\circ}$), we took the nearest
  reddening distance relationship: further details of the procedure
  are outlined in more detail by \citet{Ramsay2018}. Given that the
parallaxes are generally small (Table \ref{variables}) the distances
and hence $M_{G}$ are rather uncertain.  We selected those stars which
have positions in the ($BP-RP$)$_{o}$, $M_{G_o}$ plane which are bluer
than stars in the main sequence but with $2<M_{G_o}<10$ (i.e. brighter
than the white dwarf tracks and fainter than the BLAPs of
\citet{Pietrukowicz2017}).

There are nine stars which passed these stages and which we identified
as candidate BLAPs.  We show their sky co-ordinates, period, $g$ mag,
amplitude, $(BP-RP)$, $M_{G}$ in Table \ref{variables}.  They have a
range in period between 7.4--36.7 min, amplitudes between 0.11--0.42
mag and 16.6$<g<$19.6. We now present dedicated photometric and
spectroscopic observations of these sources which enables us to
determine their nature.

\begin{table*}
\begin{center}
\begin{tabular}{clcccccccccr}
  \hline
 ID & Name & Start Date & Start HMJD & End HMJD & No.              & Exp & Filter & $g$ & Period & Amplitude & log FAP \\
         &            &            &          & Images      &      & (s) &        & (mag)  & (min)   & (mag) & \\
  \hline
 1 & OW\,J082047.2--330107.8 & 2021-02-17 & 59262.9151 & 59263.0773 & 468  &   30. &  g  & 16.7 & 7.51 & 0.107 & --115.4 \\
   &            & 2021-02-18 & 59263.9078 & 59264.0804 & 498  &   30. &  g  & 16.7 & 7.52 & 0.092 &  --63.4 \\
   &            & 2021-02-19 & 59264.9048 & 59265.0531 & 428  &   30. &  g  & 16.7 & 7.58 & 0.072 &  --53.1 \\
   &            & 2021-02-23 & 59268.7668 & 59269.0366 & 778  &   30. &  g  & 16.7 & 7.49 & 0.066 &  --66.9 \\[5pt]
4  & OW\,J181920.9--272956.1 & 2016-07-30 & 57599.7074 & 57599.8229 & 162  &   60. &  g  & 17.4 & 15.8 & 0.186 &  --52.6 \\
   &            & 2016-08-01 & 57601.7128 & 57602.0270 & 311  &   60. &  g  & 17.4 & 15.7 & 0.186 & --129.6 \\	    
   &            & 2016-08-03 & 57603.7636 & 57604.0513 & 144	 &   60. &  g  & 17.4 & 15.7 & 0.165 &  --40.9 \\[5pt]
5  & OW\,J181100.2--273013.5 & 2016-08-04 & 57604.7208 & 57605.0370 & 215	 &  120. &  g  & 18.3 & 22.9 & 0.273 &  --69.4 \\
   &            & 2016-08-05 & 57605.7191 & 57606.0507 & 238	 &  120. &  g  & 18.3 & 22.7 & 0.210 &  --46.6 \\
   &            & 2016-08-08 & 57608.7158 & 57608.9221 & 297	 &   60. &  -  &  -   & 22.9 & 0.294 &  --71.8 \\[5pt]
8  & OW\,J182628.4--281201.9 & 2017-05-07 & 57880.9504 & 57881.0752 & 540  &   20. &  g  & 16.6 &  33.0    &  0.176     & --184.9    \\
   &            & 2017-05-09 & 57883.0261 & 57883.1926 & 720  &   20. &  g  & 16.7 &  32.9    &  0.186     &  --240.7      \\[5pt]
9  & OW\,J081530.8--342123.5 & 2021-02-20 & 59265.8426 & 59266.0604 & 628  &   30. &  g  & 16.6 & 36.8 & 0.193 & --318.8  \\
   &            & 2021-02-21 & 59266.8003 & 59267.0445 & 704  &   30. &  g  & 16.6 & 36.8 & 0.197 & --333.8  \\
   &            & 2021-02-22 & 59267.8817 & 59268.0421 & 463  &   30. &  g  & 16.6 & 36.8 & 0.202 & --239.9  \\
\hline
\end{tabular}
\end{center}
\caption{A log of photometric observations made using the 1.0m
  telescope at SAAO, Sutherland, South Africa. The last 3 columns show
  the period of the maximum peak in the Lomb Scargle periodogram; its
  amplitude (peak-to-peak) and its False Alarm Probability.}
\label{photfollowup}
\end{table*}

\begin{table*}
\begin{center}
\begin{tabular}{clcccccccc}
  \hline
ID & Name & Date & Telescope & Instrument & Grating & Slit Width & Pos.Ang & Gr.Ang & Exposure \\ 
   &      &      &           &            &         & (arcsec) & (deg) & (deg) &  (sec) \\
  \hline
1 & OW\,J082047.2--330107.8 & 2016-02-10 & SALT &  RSS  & PG0900  & 1.5 &   54 & 14.38 & 700 \\
  &          & 2017-11-30 & SALT &  RSS  & PG0900  & 1.5 & --101 & 14.38 & 804 \\[5pt]
3 & OW\,J181227.9--293849.3 & 2016-09-30 & SALT &  RSS  & PG2300  & 1.5 &   45 & 30.88 & 2$\times$600  \\[5pt]
4 & OW\,J181920.9--272956.1 & 2016-08-05 & WHT  & ISIS  & 158R/300B & 1.0 & 10.8 & 58.8/58.3 & 30$\times$90 \\
  &         & 2016-09-07 & SALT &  RSS  & PG2300  & 2.0 & 56 & 30.88 & 15$\times$60 \\[5pt]
5 & OW\,J181100.2--273013.5 & 2016-05-15 & SALT &  RSS  & PG0900  & 1.5 & 291 & 14.38 & 1800  \\
  &          & 2016-10-22 & SALT &  RSS  & PG2300  & 1.5 & 0   & 30.88 & 720  \\[5pt]
6 & OW\,J181038.5--251608.6 & 2016-05-10 & SALT &  RSS  & PG0900  & 1.5 & 233 & 14.38  & 2200 \\[5pt]
7 & OW\,J175848.2--271653.6 & 2016-06-28 & SALT &  RSS  & PG0900  & 1.5 & 177 & 14.38 & 50$\times$30 \\
  &           & 2016-07-09 & SALT &  RSS  & PG0900  & 1.5 & 177 & 14.38 & 50$\times$30\\[5pt]
9 & OW\,J081530.8--342123.5 & 2017-11-18 & SALT &  RSS  & PG0900  & 1.5 & --99 & 14.38 & 840 \\
  &                       & 2021-05-01 & SALT & RSS & PG0900 & 1.5 & 98 & 14.38 & 6$\times$400 \\
  &                       & 2021-05-02 & SALT & RSS & PG0900 & 1.5 & 99&  14.38 & 6$\times$400 \\
  &                       & 2021-05-09 & SALT & RSS & PG0900 & 1.5 & 99 & 14.38 & 6$\times$400 \\
      \hline
\end{tabular}
\end{center}
\caption{A log of spectroscopic observations showing the
  date; telescope; instrument; grating; slit-width; position angle; grating angle and exposure time.}
\label{specfollowup}
\end{table*}

\section{High speed photometry}

We obtained follow-up photometry of five of the sources outlined in
Table \ref{variables} using the SAAO 1.0m telescope and the Sutherland
High-speed Optical Cameras (SHOC; \citet{Coppejans2013}) at
Sutherland, South Africa.  We indicate the time and duration of these
photometric observations together with the exposure time and filter in
Table \ref{photfollowup}.

The data were reduced using standard procedures in IRAF. Light curves
were derived using aperture-corrected photometry, calibrated using
Pan-STARRS $g$-band photometry \citep{Chambers2016}.  We then used the
{\tt VARTOOLS} suite of software \citep{Hartman2008} to obtain a power
spectrum for each light curve using the LS implementation. We show the
period, amplitude and FAP of the peak in the power spectra in Table
\ref{photfollowup}.

\section{Follow-up spectroscopic Observations}

We obtained follow-up spectra of six of our candidate BLAPs using the
William Herschel Telescope (WHT) and the Southern African Large
Telescope (SALT). Our main aim was to detect principal line features
which could be used for identification purposes and possibly to
constrain spectral parameters.

The data were reduced using standard procedures (see
\citet{Macfarlane2017a} for more details) and an overview of the
observations are given in Table \ref{specfollowup}.  {The bulk of the
  spectra were obtained using the Robert Stobie Spectrograph (RSS;
  \citet{Kobulnicky2003}) on SALT, using either the PG0900 or the
  PG2300 grating. The former covered a wavelength range of $\sim$
  3980-6980 {\AA} with a dispersion of 0.97 {\AA}, whereas the latter
  covered the wavelength range of $\sim$ 3900-4950 {\AA} with a
  dispersion of 0.34 {\AA}.}

\begin{figure*}
  \centering
  \hspace{1cm} \includegraphics[width=16cm]{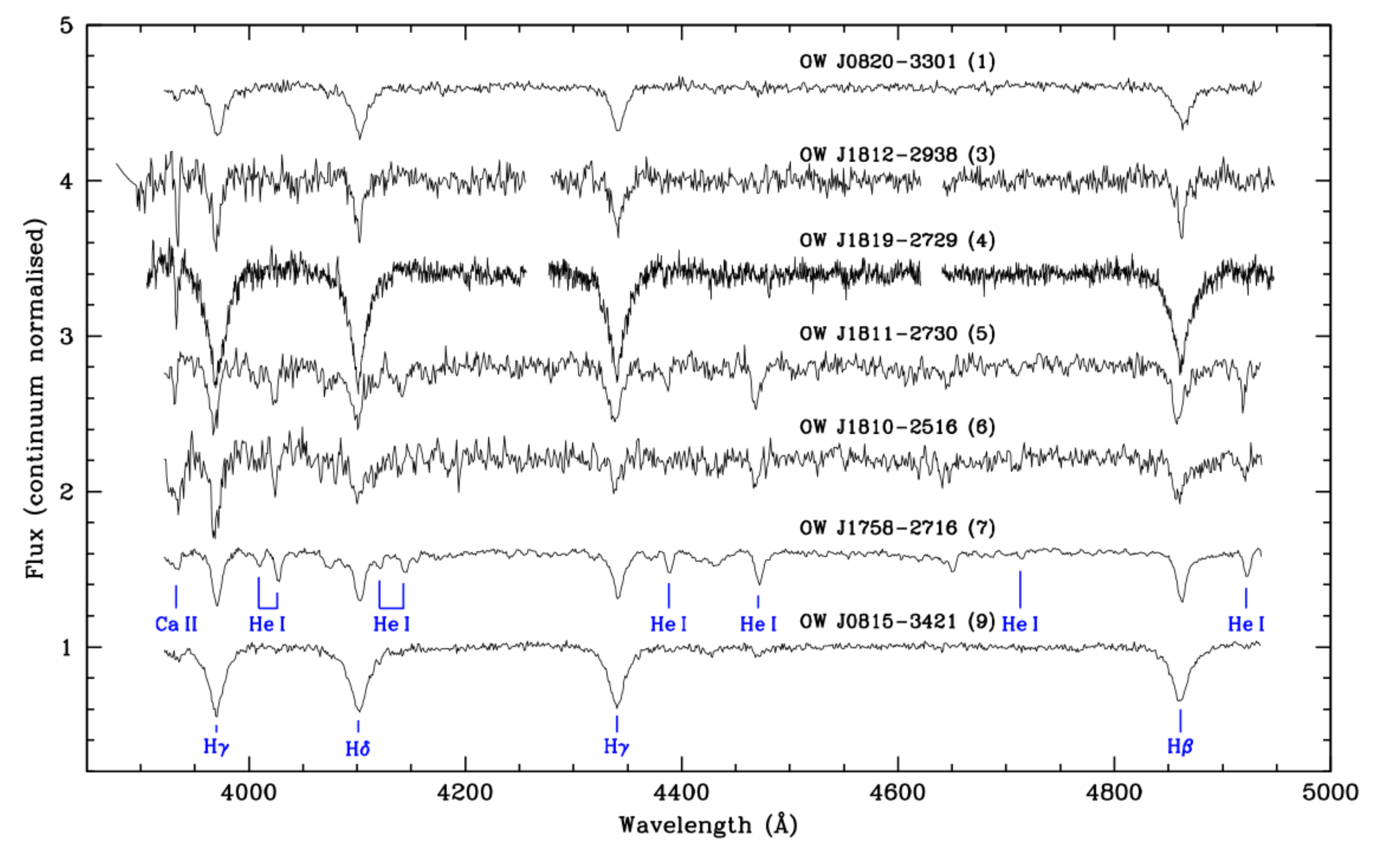}\hfill
  \caption{The continuum normalised SALT spectra of our targets where
    each spectrum has been shifted by +0.5 flux units. We note the
    wavelength of key lines in the spectra.}
\label{specBLAPcont}
\end{figure*}

All objects show Balmer lines in absorption with some showing helium
lines (see Fig.\,\ref{specBLAPcont}). Atmospheric parameters such as
effective temperature, \teff, surface gravity, \logg, helium
abundance, $\log{y}=\log\frac{n(He)}{n(H)}$ were determined by fitting
the rest-wavelength corrected Balmer and helium lines using the
\texttt{FITSB2} routine \citep{napiwotzki2004}.  Two grids of model
atmospheres were used, which cover wide ranges in parameter
space. Both model grids were constructed by combining LTE calculation
of the atmospheric stratification with the \textsc{atlas12} code
\citep{1996ASPC..108..160K}, including full line-blanketing, and of
NLTE occupation densities for extensive hydrogen and helium model
atoms with the \textsc{detail} code \citep{1981PhDT}. The emergent
optical hydrogen and helium lines spectrum was finally synthesised
with the \textsc{surface} code \citep{Butler_1985,2011JPhCS.328a2015P}
using state-of-the line broadening tables. Recent updates and
extensions of the codes are described in
\citet{2018A&A...615L...5I}. \textsc{atlas12} allows individual
non-solar abundance pattern to be implemented. The first grid of
models uses an abundance pattern typical for sdB stars
\citep{naslim13} \citep[additional details can be found
  in][]{schaffenroth2021}. The grid covers the temperature range
\teff=9,000K to 55,000K, surface gravities from \logg=4.6 to 6.6 and
helium to hydrogen ratio y from 10$^{-4}$ to 10 times that of the sun.

Another grid was provided by H\"ammerich (priv. comm.) with models
covering helium to hydrogen abundance ratio from 10$^{-4}$ to 5 times
that of the sun, but using the solar metal abundance pattern, which
allowed the range for the surfaces gravities to be extended to
\logg=4.0, which was necessary for three program stars.

\section{Individual objects}

\begin{figure}
  \centering
  \hspace{0.5cm} \includegraphics[width=8.2cm]{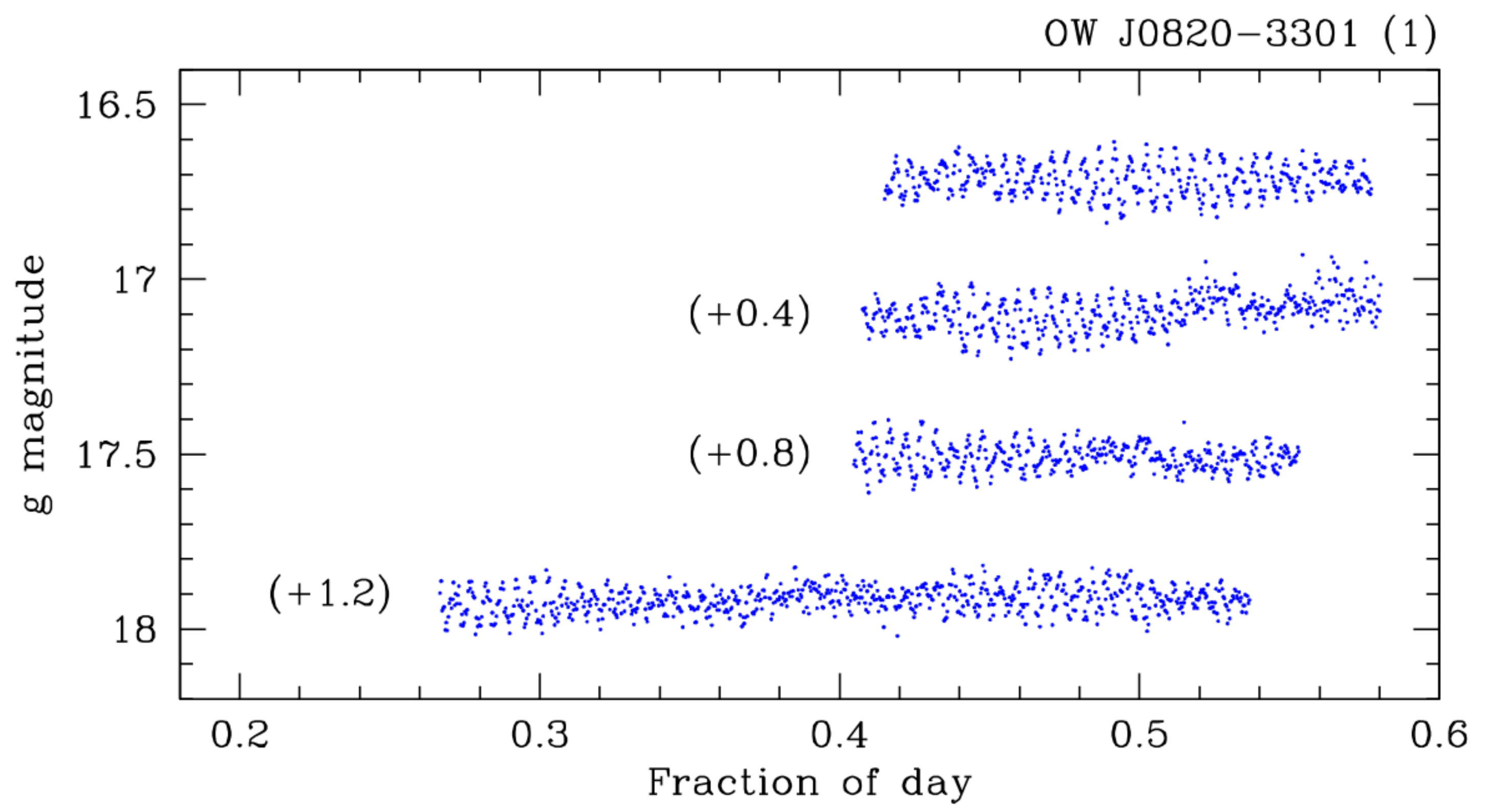}\hfill
  \caption{Light curves of OW\,J0820--3301 obtained with the SAAO 1-m
    telescope. The top light curve is shown at the correct brightness,
    whereas subsequent light curves (top-to-bottom) are displaced for
    display purposes only.}
\label{lcOWJ0820}
\end{figure}

\subsection{OW\,J082047.2--330107.8}

\begin{figure}
\centering
\hspace{0.5cm} \includegraphics[width=8cm]{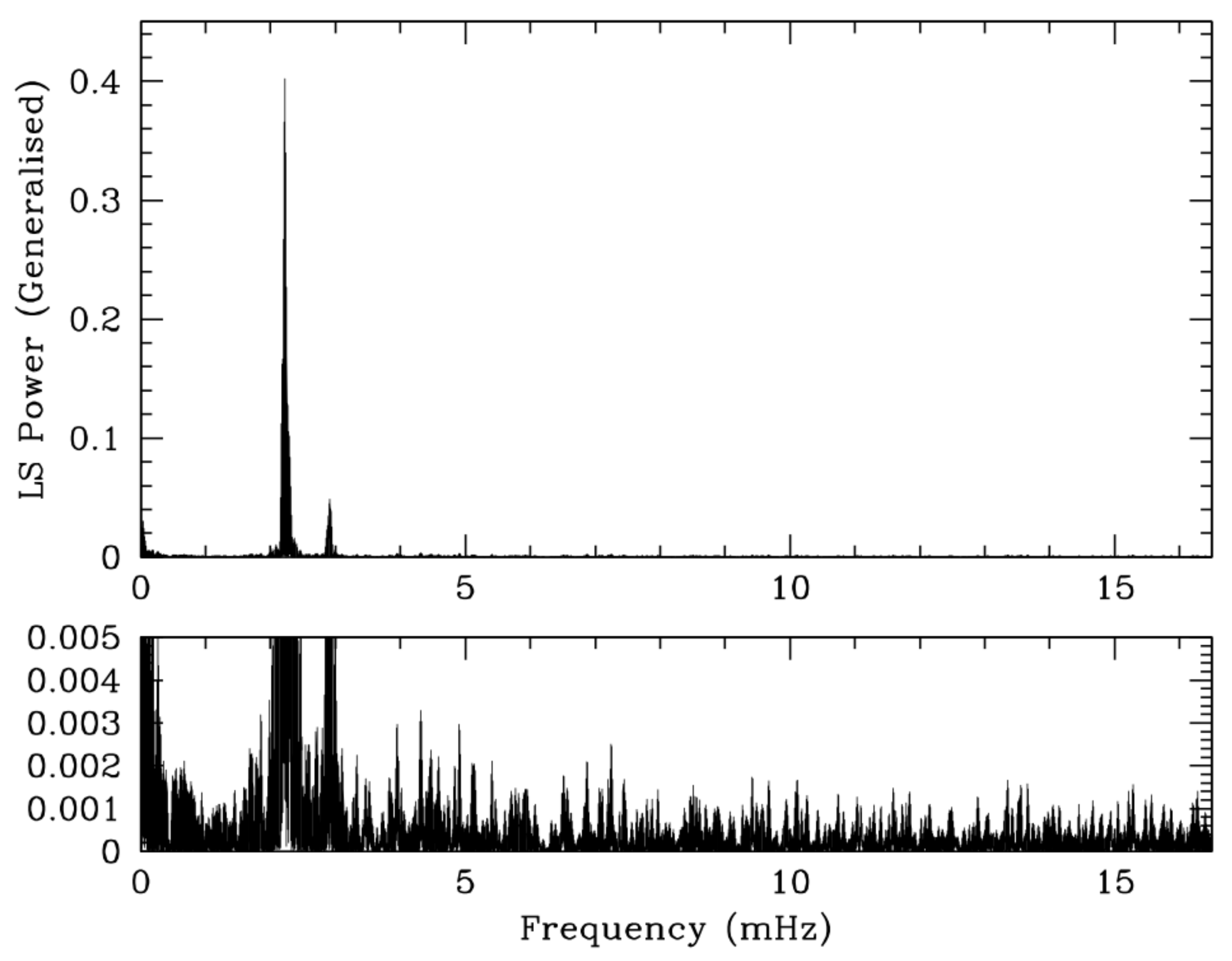}
\caption{Generalised Lomb-Scargle power spectrum of OW\,J0820-3301.}
\label{lsblap1}
\end{figure}

The OW photometric observations of OW\,J082047.2--330107.8 (ID 1 in
Table \ref{variables}, and hereafter referred to as OW\,J0820--3301)
showed evidence for periodic modulation on a period of 7.4 min and
0.13 mag in amplitude. We obtained followup photometry on four nights
in Feb 2021 (Table \ref{photfollowup}) with a significant periodic
signal on a period of 7.5 min on each occasion. The amplitude appears
to be variable as does the significance of the period. In
Fig. \ref{lcOWJ0820} we show these light curves: the amplitude of the
modulation does change over time. An LS power spectrum
(Fig. \ref{lsblap1}) shows an additional, less prominent, peak
corresponding to a shorter period (5.7 min).

The multi-periodic nature of OW\,J0820--3301, and its high temperature
(Table \ref{spec-fits}) are consistent with the properties of
V361\,Hya (or EC\,14026) variables
\citep{Kilkenny1997,Heber2016,LynasGray2021}.  Pulsation amplitudes in
V361\,Hya variables are typically less than 1\% compared with nearly
10\% in OW\,J0820--3301.  However, several other V361\,Hya variables
have been observed with significant semi-amplitudes, including
PG\,1605+072 ($P$ = 8.0\,min, $a\sim0.064$\,mag) \citep{Koen1998},
Balloon\,090100001 ($P$ = 5.9\,min, $a\sim0.060$\,mag)
\citep{Baran2009} and CS\,1246 ($P$ = 6.2\,min, $a\sim0.025$\,mag)
\citep{Barlow2010}.

Interestingly, the large amplitude oscillation observed in CS\,1246
decayed to $a<0.005$\,mag in the following eight years
\citep{Hutchens2017}, while amplitude variations appear to be
important in other V361\,Hya variables \citep{Reed2004,Kilkenny2010}.

\begin{figure}
\includegraphics[angle=0,width=8cm]{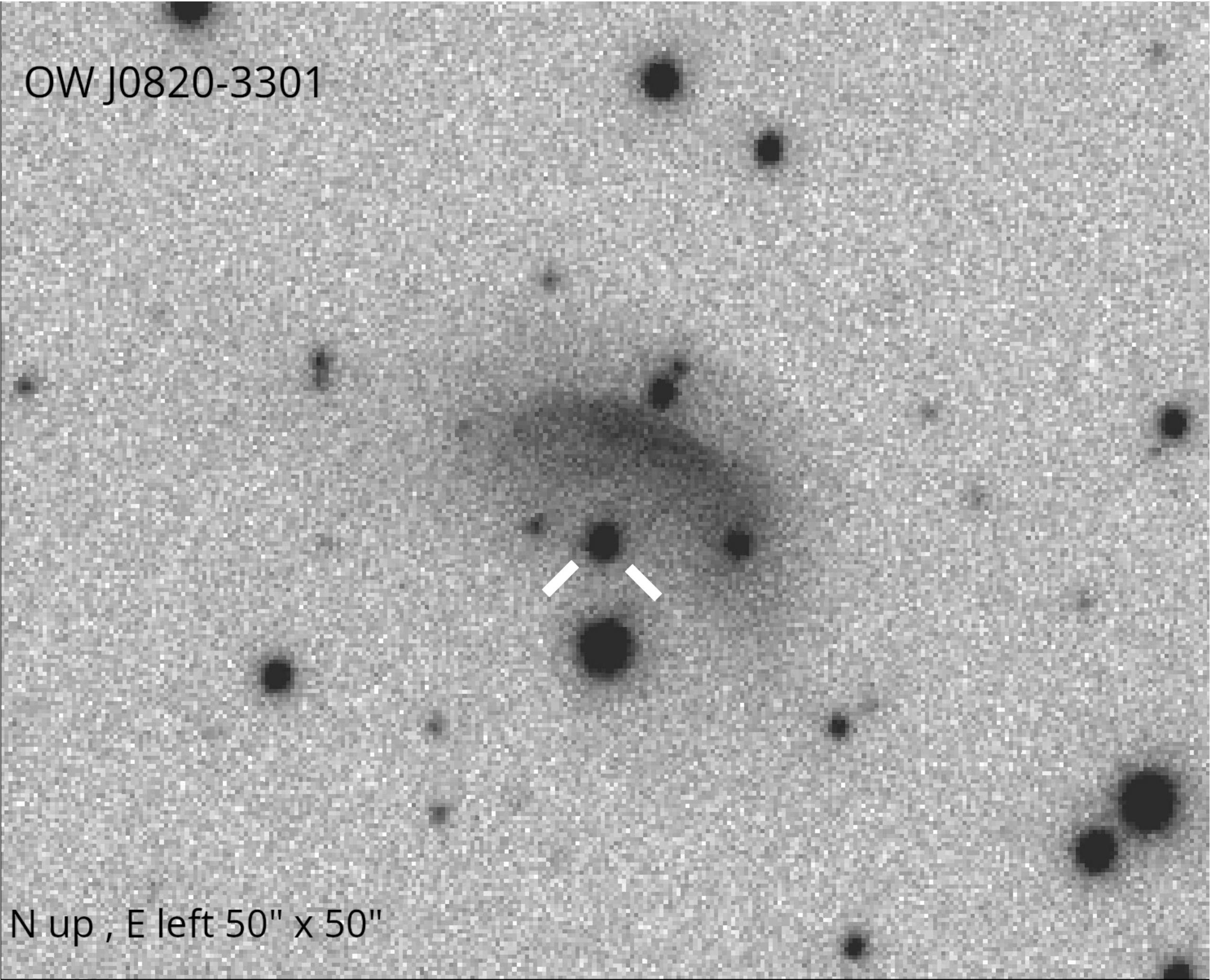}\hfill
\caption{An image of the immediate field of OW\,J0820--3301 in
  H$\alpha$, combining three exposures taken during the VPHAS+ survey
  \citep{Drew2014}. It is 50$^{\prime\prime} \times50^{\prime\prime}$
  wide and the target is in the center, indicated by the white
  dashes. An extended nebula lies to the north of OW J0820-3301.}
\label{nebula}
\end{figure}

The SALT spectrum of OW\,J0820--3301 obtained in 2016 (see details in
Table \ref{specfollowup}) showed a nebular spectrum of a Str\"omgren
sphere superimposed on broader Balmer absorption lines.  After careful
sky subtraction, the emission line spectrum was separated from the
stellar spectrum. The latter is shown in
Fig. \ref{specBLAPcont}. Fitting this spectrum indicates
OW\,J0820--3301 has a temperature of \teff$\sim37,400$K and \logg
$\sim$5.7 (Table \ref{spec-fits}).

We performed a search for extended emission around our targets using
the AAO/UKST
H$\alpha$\footnote{\url{http://www-wfau.roe.ac.uk/sss/halpha}} and
VPHAS+ \citep{Drew2014} surveys and extracted an image for all
sources. We found evidence for extended emission from around
OW\,J0820--3301: in Fig. \ref{nebula} we show an image taken in
H$\alpha$ extracted from the VPHAS+ survey \citep{Drew2014} which has
higher resolution. The DECaPS
survey \footnote{\url{http://decaps.skymaps.info}} shows the nebula as
prominently in green colour, indicative of strong O{\sc iii} or
H$\beta$ emission. Without more detailed spectroscopic observations
across the nebula (and ideally integral field unit observations) we
are unable to be certain that the nebula is associated with
OW\,J0820--3301. However, using statistical arguments we can determine
the likelihood that two 'interesting' objects are located within (say)
2$^{\prime\prime}$ of each other. The Gaia EDR3 catalogue
\citep{Gaia2021} shows 56 stars within 1$^{\prime}$ radius of
OW\,J0820--3301 which are brighter than $G<19$ mag (85 for $G<$20
mag). The probability that this is a chance alignment is 0.03 percent
($G<19$) and 0.07 percent.

{\it If} OW\,J0820--3301 and the nebula are associated then we can
estimate the age of the nebula. The Gaia EDR3 data \citep{Gaia2021}
has a best fit distance to OW J0820--3301 of 3.7 kpc. The H$\alpha$
image shows a nebula which has a diameter along its semi-major axis of
$\sim8^{\prime\prime}$. For a distance of 3.7 kpc, simple geometry
indicates a semi-major axis of $\sim$0.14 pc. If we assume a velocity
of the expanding material to be 10--40 km s$^{-1}$
\citep{CorradiSchwarz1995} we find an age of the nebula of
$\sim$3400--13600 years. More dedicated spectroscopic observations of
the nebula are required.

\subsection{OW\,J181227.9--293848.3}

We identified OW\,J181227.9--293848.3 (ID 3 in Table \ref{variables},
and hereafter referred to as OW\,J1812--2938) as having a very
significant peak in its power spectrum at 10.8 min and an amplitude of
0.28 mag. No followup photometry was obtained of this source but we
did obtain two spectra using SALT in 2016 (Table
\ref{specfollowup}). We show the combined normalised spectrum in
Fig. \ref{specBLAPcont} which shows Balmer lines in absorption and a
prominent Ca II absorption line.

Fitting this spectrum indicates OW\,J1812--2938 has a temperature
\teff$\sim31,000$K and \logg $\sim$4.7 (Table \ref{spec-fits}). These
results, together with its period and amplitude, place it in the
parameter range consistent with it being a BLAP, and we therefore
designate it OW-BLAP-1.

\subsection{OW\,J181920.9--272956.3}

\begin{figure}
  \centering
  \hspace{0.5cm} \includegraphics[width=8.2cm]{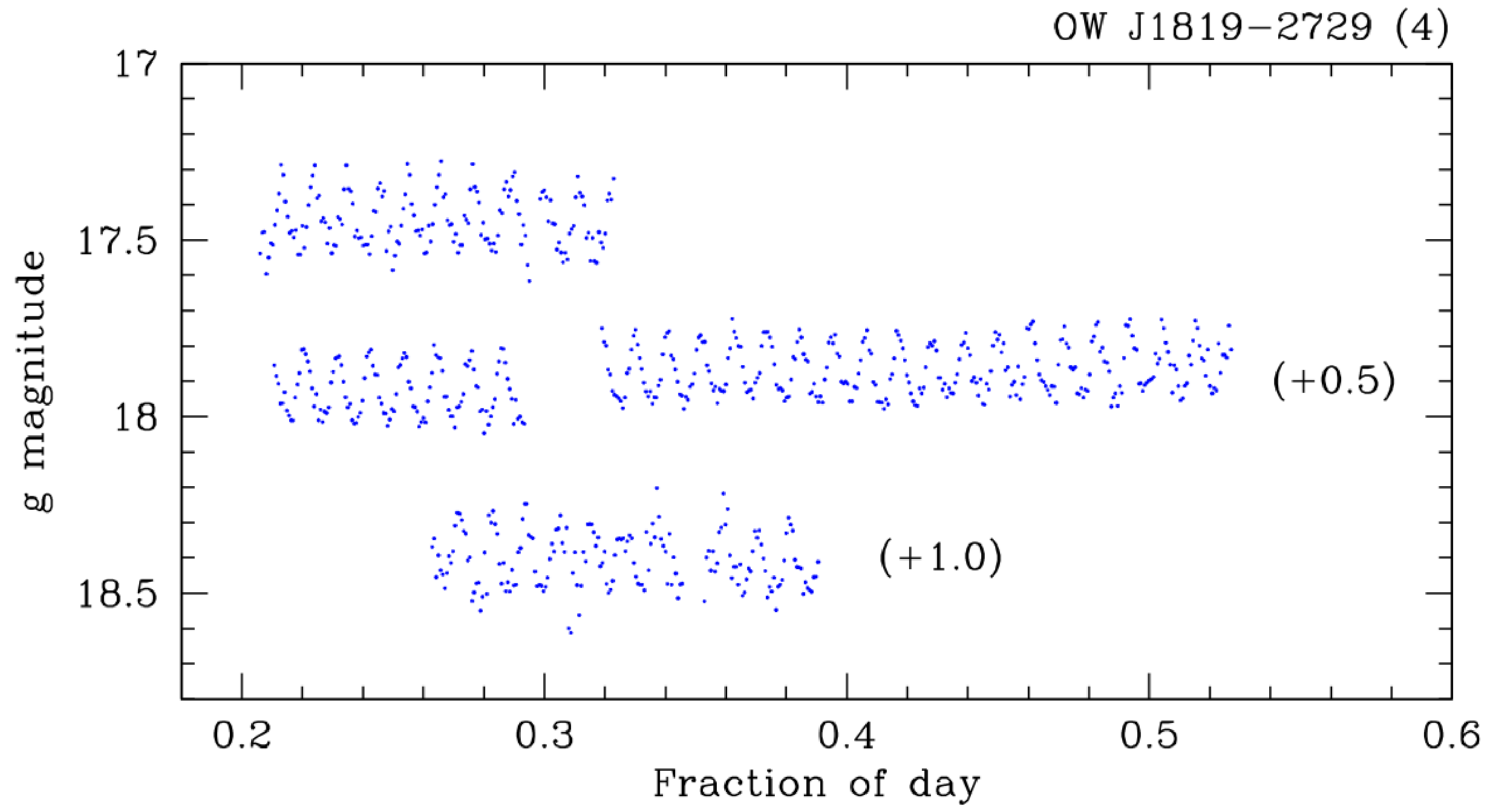}\hfill
  \caption{Light curves of OW\,J1819--2729 obtained with the SAAO 1-m
    telescope.}
\label{lcOWJ1819}
\end{figure}

We identified OW\,J181920.9--272956.3 (ID 4 in Table \ref{variables},
and hereafter referred to as OW\,J1819--2729) as having a very
significant peak in its power spectrum at 15.7 min and an amplitude of
0.19 mag. We obtained followup photometry of this source using the
SAAO 1.0m telescope on three nights with multiple 60 sec exposures on
each occasion (c.f. Table \ref{photfollowup}). The light curves are
shown in Fig. ~\ref{lcOWJ1819}. These data confirm the 15.9 min period
found from OW data although there is also a peak in the power spectra
at half the 15.9 min period. There are some side-lobes of the main
peak on the night of 2016-08-03 but this has the smallest number of
photometric points. The folded light curves are sinusodial and
symmetric.

We obtained spectra of OW\,J1819--2729 using SALT, SAAO 1.9m
telescope and WHT. The longest series was obtained using the WHT where
a series of 30 low resolution spectra were obtained. They were
obtained at relatively high airmass (1.8--2.0) and the individual
spectra were of low signal-to-noise ratio. However, the co-added spectra
revealed Balmer absorption lines, with only marginal evidence
for helium lines. The ratio of the equivalent width (Ca\,{\sc ii} H+K,
H${\epsilon}$) is consistent with an A-type spectrum. 

The temperature (\teff$\sim$12,300 K) derived from the spectral fits
(Table \ref{spec-fits}) is the lowest of the stars in our sample and
is similar to the pulsating sdA stars \citep{Bell2018}.  However,
there is only one star in the \citet{Bell2018} set of new stars which
has a period (4.6 min) which is even remotely close to that seen here
but the amplitude of the pulsation of that star is $<$0.4
percent. These observations indicate that OW\,J1819--2729 is a
possible sdAV variable star.

\subsection{OW\,J181100.2--273013.3}

\begin{figure}
  \centering
  \hspace{0.5cm} \includegraphics[width=8.2cm]{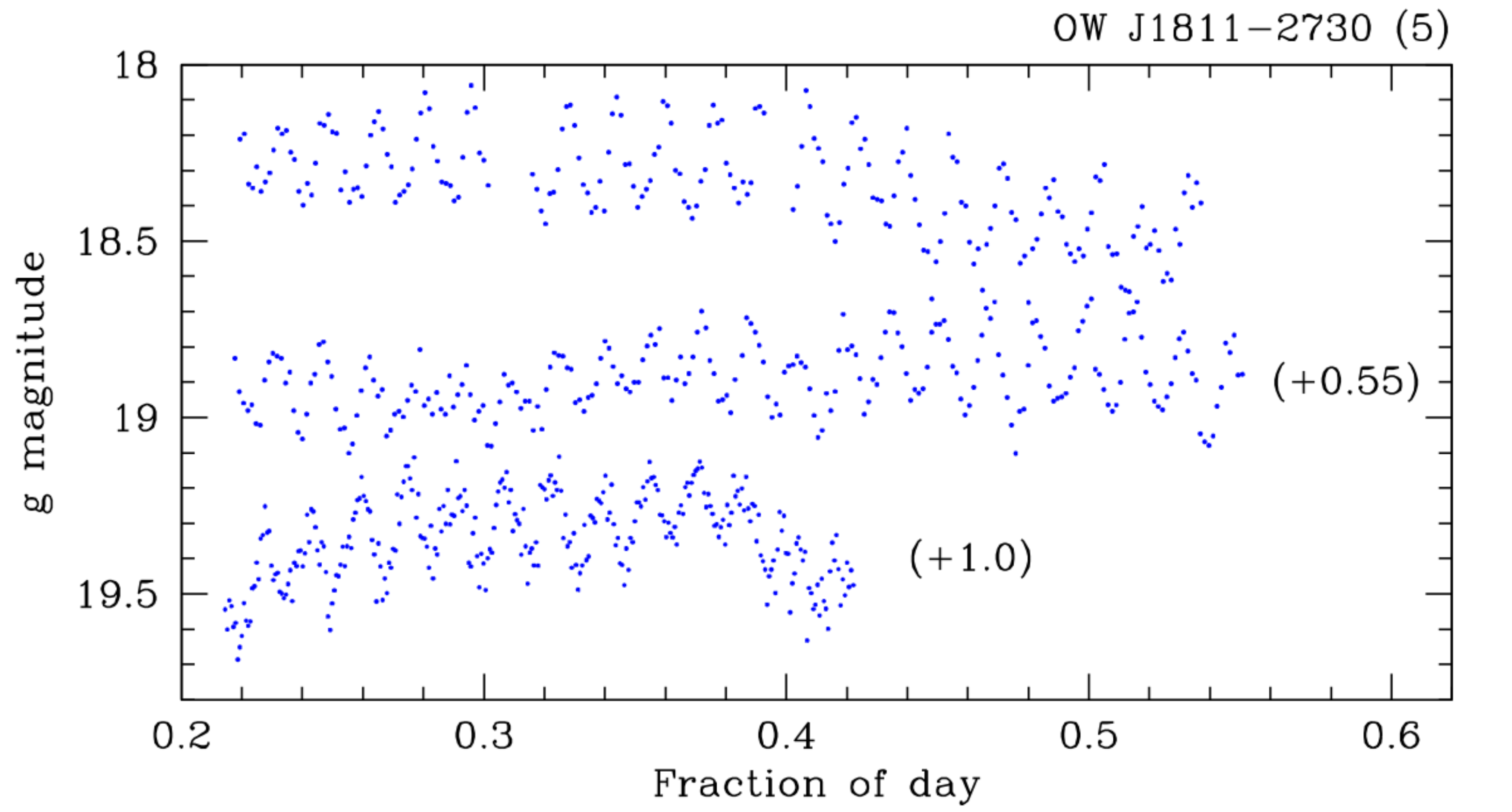}\hfill
  \caption{Light curves of OW\,J1811--2730 obtained with the SAAO 1-m
    telescope.}
\label{lcOWJ1811}
\end{figure}

\begin{figure}
  \centering
  \includegraphics[width=8.5cm]{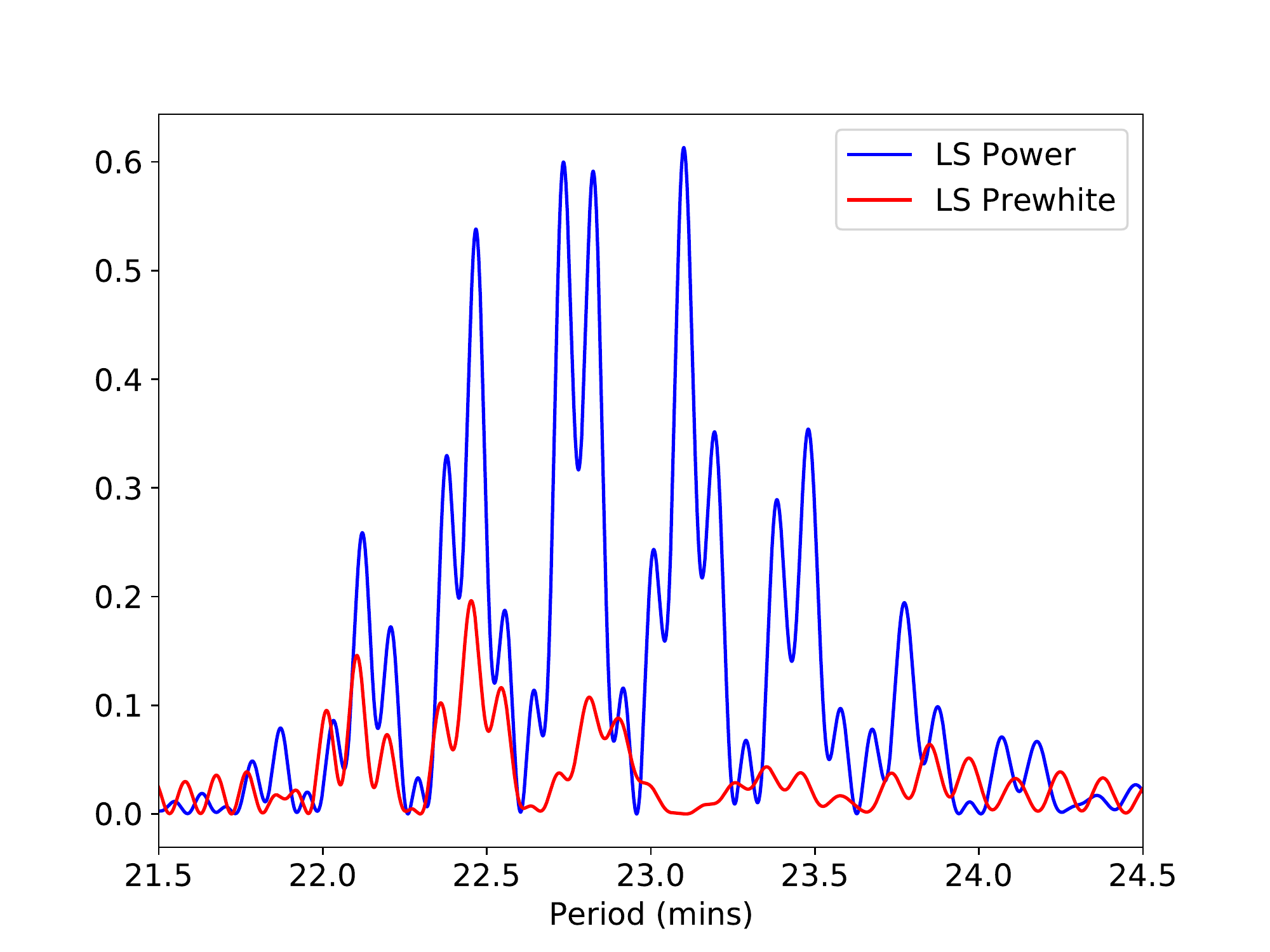}\hfill
  \caption{The Lomb-Scargle power spectrum of the combined light curve
    of OW\,J1811--2730 (after global trends have been removed) and the
    Lomb-Scargle power spectrum after the light curve has been
    pre-whitened by removing the period with maximum power.}
\label{lcOWJ1811prew}
\end{figure}

OW\,J181100.2--273013.3 (ID 5 in Table \ref{variables}, and hereafter
referred to as OW\,J1811--2730) was observed on three occasions using
the SAAO 1.0m telescope: at two epochs the exposure times were 120 sec
and at one epoch 60 sec (Table \ref{photfollowup}). The light curves
from the individual nights are shown in Fig. \ref{lcOWJ1811}.

The period derived from each of the three nights is 22.7 -- 22.9 min
with an amplitude in the range 0.21--0.29 mag which is fully
consistent with the 23.0 min and 0.21 mag from the OW survey data. To
search for other periods in the light curve, we removed global trends
in each night (which may have been due to differential colour terms
between the comparison stars) and obtained a LS power spectrum
(Fig. \ref{lcOWJ1811prew}). There is evidence for the peaks in the
power spectrum being split, but these largely disappear after
pre-whitening on the period with peak power. We conclude there is
little evidence for a second period in this source.

The mean spectrum shown in Fig. \ref{specBLAPcont} shows the Balmer
lines in absorption together with He\,{\sc i} also in absorption. The
spectral fits (Table \ref{spec-fits}) indicates OW\,J1811--2730 has a
temperature $T_{\rm eff}\sim$27,000 K and log $g\sim$4.8. Together
with the photometric properties indicates that OW\,J1811--2730 is a
BLAP and we designate OW-BLAP2.

\subsection{OW\,J181038.5--251608.6}

We identified OW\,J181038.5--251608.6 (ID 6 in Table \ref{variables},
and hereafter referred to as OW\,J1810--2516) as having a very
significant peak in its power spectrum at 28.9 min and an amplitude of
0.42 mag. At $g$=19.6 mag it is the faintest target in our sample. We
obtained one spectrum of OW\,J1810--2516 using SALT and the RSS and
its normalised spectrum is shown in Fig. \ref{specBLAPcont}. Balmer
lines are seen in absorption as are several lines due to helium.  The
fits to the spectra (Table \ref{spec-fits}) indicate a temperature
$T_{\rm eff}\sim30,000$K and \logg $\sim$4.2. Together with the period
and amplitude we identify this as a BLAP and name it OW-BLAP-3.

\subsection{OW\,J175848.2--271653.6}

\begin{figure}
\centering
\hspace{1cm} \includegraphics[width=8cm]{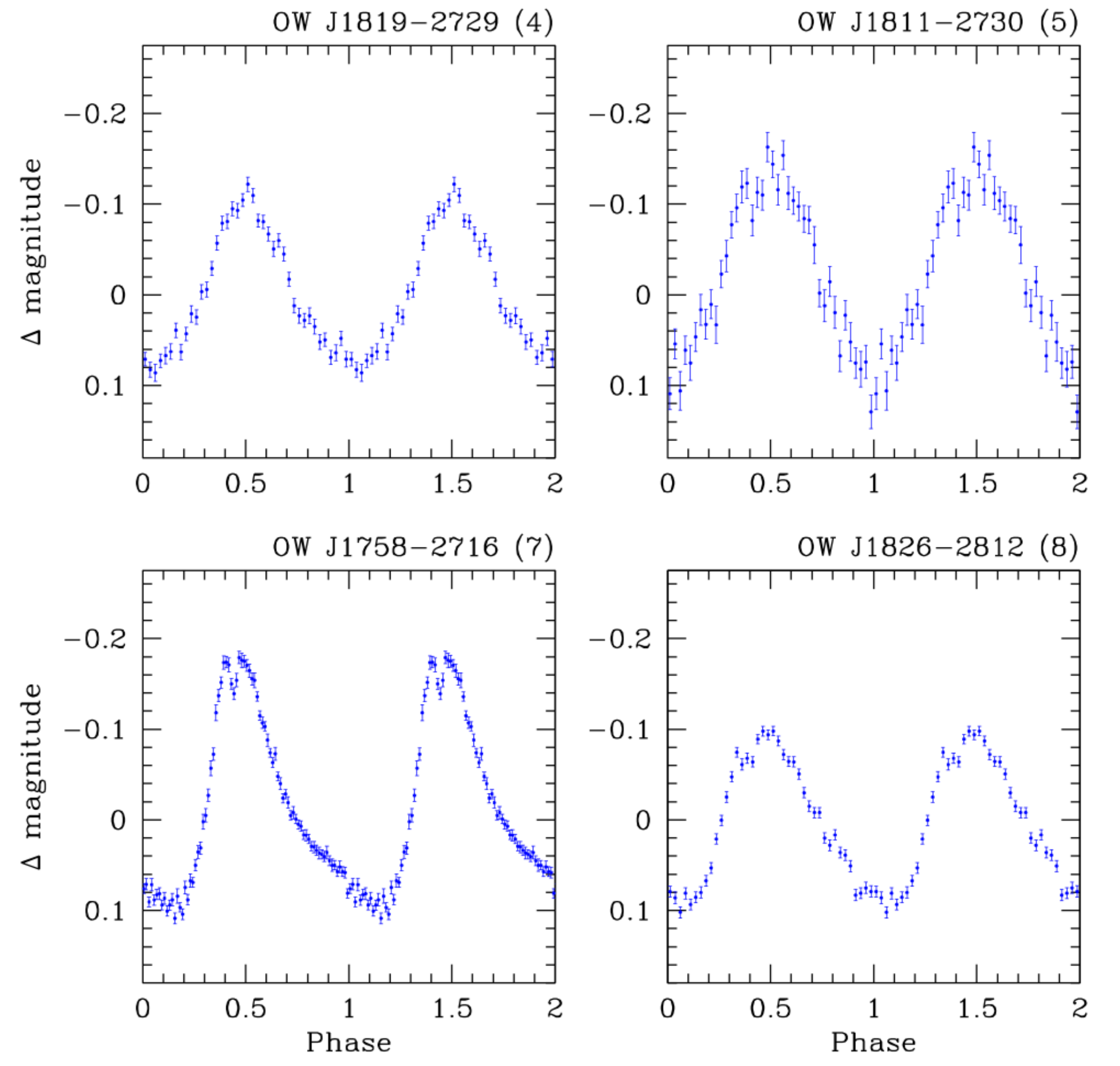}\hfill
\caption{The phased and averaged light curves of OW\,J1819--2729 ($P$
  = 15.715 min, top left), OW\,J1811--2730 ($P$ = 22.915 min, top
  right), OW\,J1758--2716 ($P$ = 32.00 min, bottom left) and
  OW\,J1826--2812 ($P$ = 32.960 min, bottom right).}
\label{avphot}
\end{figure}

OW\,J175848.2--271653.6 (ID 7 in Table \ref{variables},
OW\,J1758--2716) was reported as a variable with a period of 32.5 min
and amplitude of 0.25 mag \citep{Macfarlane2017a}. The folded light
curve indicated the source was similar to that of $\delta$ Sct
stars. However, an optical spectrum indicated the star was more
similar to a B spectral type than an A-F spectral type which would be
expected from a $\delta$ Sct star. Intriguingly the phase folded light
curve showed a curious `notch' at peak brightness: we display the data
reported in \citet{Macfarlane2017a} folded on the main period in
Fig. \ref{avphot}. \citet{Macfarlane2017a} made the comparison between
the notch observed here with the `Bump Cepheids' \citep{Mihalas2003}.
We obtained an additional set of 50 spectra taken on two nights in
2016 (see Table \ref{specfollowup}). Fitting the mean spectrum we find
$T_{\rm eff}\sim27,000$K and log g$\sim$4.2 (Table \ref{spec-fits})
and conclude that OW\,J1758--2716 is a BLAP and name it OW-BLAP-4.

\subsection{OW\,J182628.4--2812019}

\begin{figure}
\centering
\includegraphics[width=8.2cm]{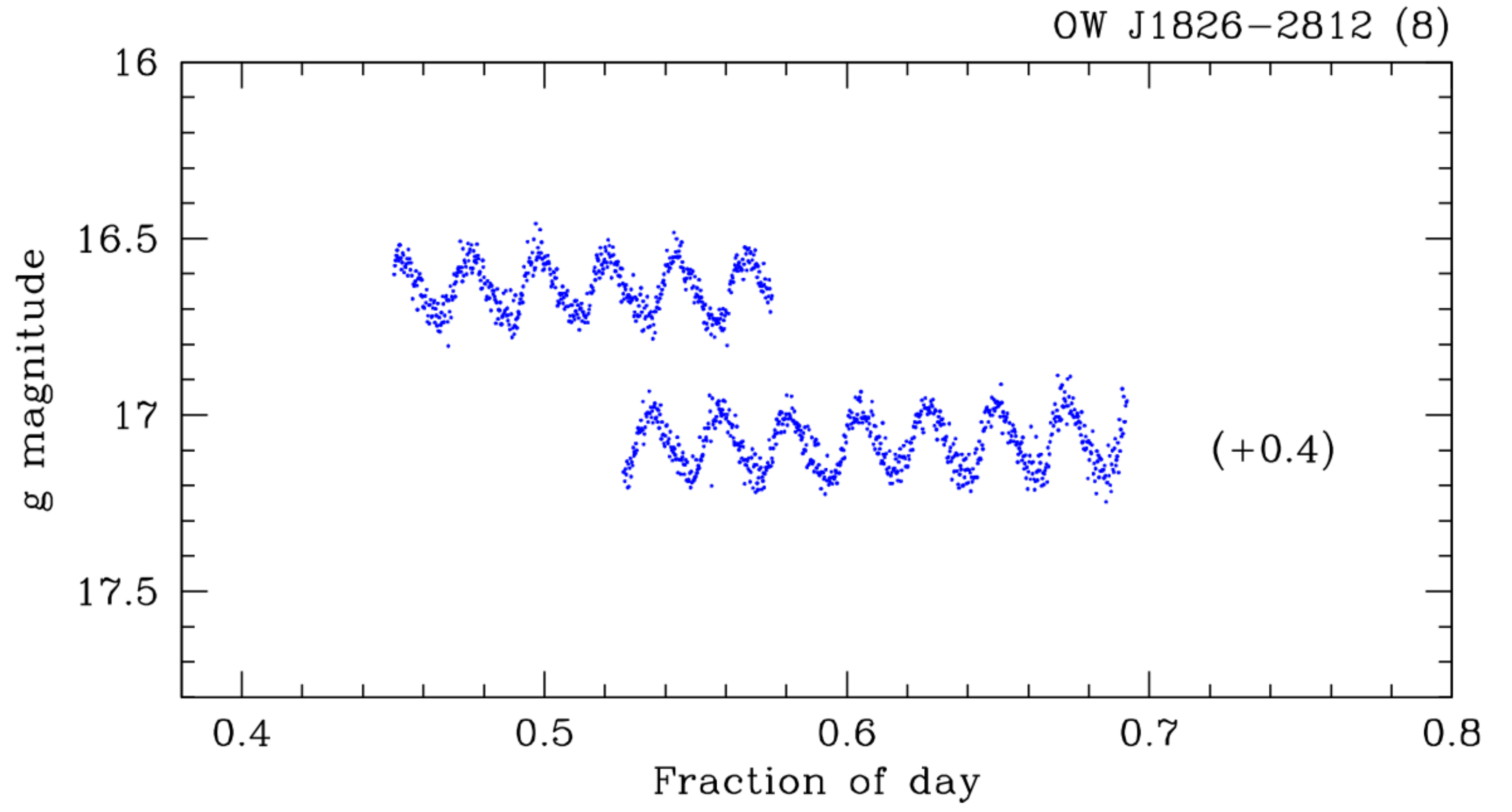}\hfill
\caption{Light curves of OW J1826-2812 obtained using the SAAO 1-m telescope.}
\label{lcOWJ1826}
\end{figure}

OW\,J182628.4--2812019 (ID 8 in Table \ref{variables},
OW\,J1826--2812) was observed on two nights in May 2017 (Table
\ref{photfollowup}) using the 1.0m telescope and SHOC. The period
which we determine (32.9--33.0 min) is slightly longer than found
using the original OW data (32.4 min) and with a higher amplitude
($\sim$0.18 mag compared to 0.12 mag). We show the light curves in
Fig. \ref{lcOWJ1826} and the phase folded binned light curve in
Fig. \ref{avphot}. There is a curious `ledge' like feature just before
maximum light. We return to this in \S \ref{discussion}. Spectroscopic
observations are still required to determine the nature of this
source.

\subsection{OW\,J081530.8-342123.5}

\begin{figure}
  \centering
  \hspace{0.5cm} \includegraphics[width=8.2cm]{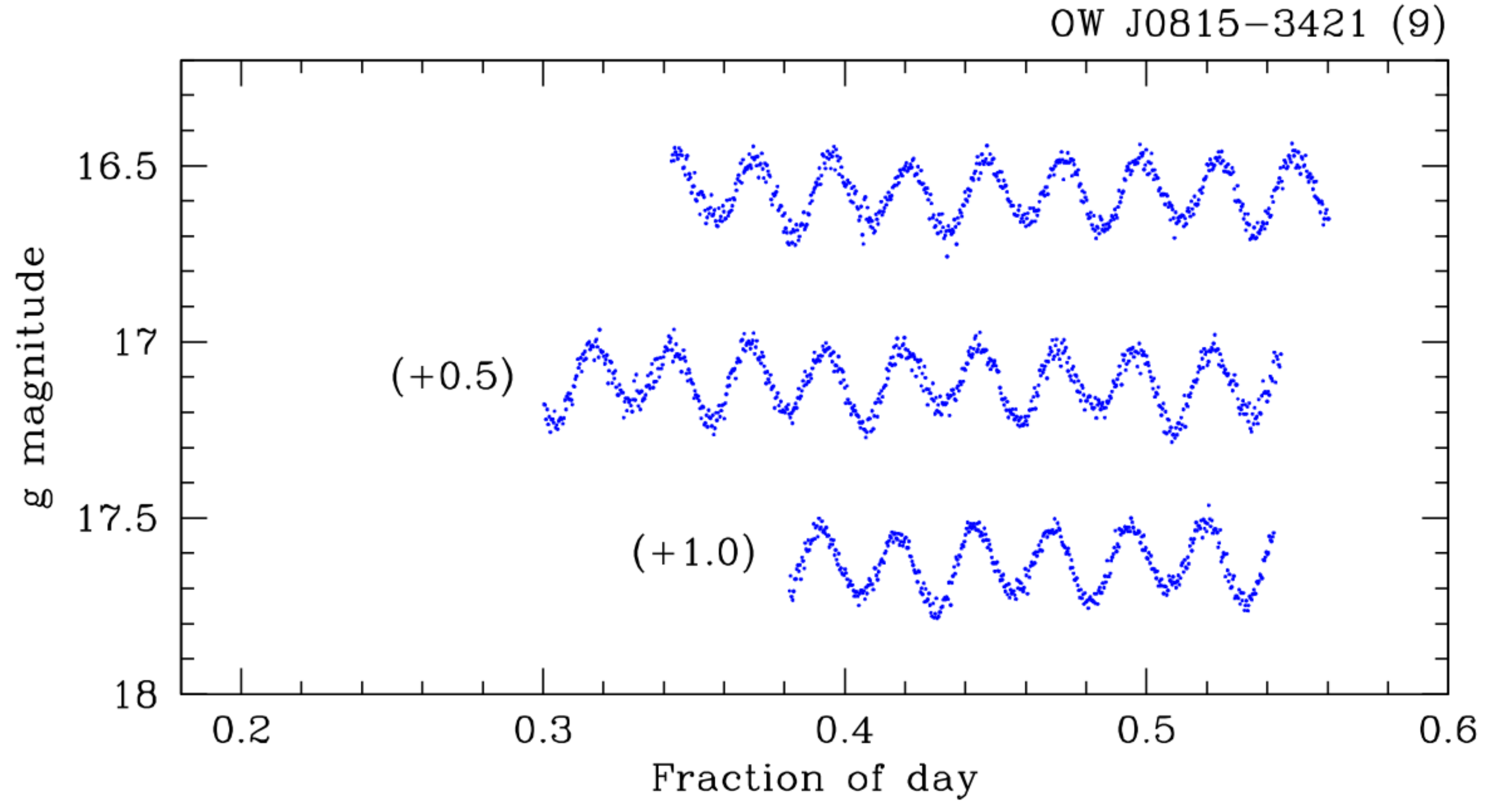}\hfill
  \caption{Light curves of OW\,J0815--3421 obtained with the SAAO 1-m telescope.}
\label{lcOWJ0815}
\end{figure}

OW\,J081530.8--342123.5 (ID 9 in Table \ref{variables}, and hereafter
referred to as OW\,J0815--3421) has the longest apparent period of the
candidate BLAPs shown in Table 1 (36.7 min) and an amplitude of 0.20
mag. Follow-up photometric observations were obtained on three nights
in Feb 2021 (Table \ref{photfollowup}). A strong periodic signal was
seen in the light curves with the LS periodogram indicating a period
of 36.8 min which is consistent with the OW period. However, in
Fig. \ref{lcOWJ0815} we show the individual light curves from these
nights: there is evidence that the depth of every second minimum is
deeper than the preceding minimum. This suggests that the true period,
possibly an orbital ellipsoidal modulation, is 73.6 min.

We obtained a single spectrum of OW\,J0815--3421 in 2017 (Table
\ref{specfollowup}) which showed the Balmer series in absorption
(Fig. \ref{specBLAPcont}). A fit to this spectrum indicates a
temperature and gravity consistent with that of an sdB star. To search
for radial velocity variations we obtained a series of spectra using
SALT in May 2021. This revealed a clear radial velocity variation (a
semi-amplitude of 376$\pm$15 km s$^{-1}$) on a period of 73.6 min: we
show the phase folded radial velocity curve in
Fig. \ref{OWJ0815-spec-rv-fit}.  A fit to the combined spectrum (made
after accounting for the radial velocity variation of the individual
spectra) indicates a temperature $T_\mathrm{eff}$ = $26,500\pm1000$ K
and $\log g$=$5.68\pm0.15$.

We can now use this information to model and constrain the resulting
stellar parameters using the photometric light curve using
\textit{Lcurve}\footnote{\url{https://github.com/trmrsh/cpp-lcurve}},
(see \citet{Copperwheat2011} for a detailed description). To model the
system, we use a spherical star for the white dwarf and use a Roche
geometry for the sdB companion. The free parameters in the model are
the period ($P$), time of inferior conjunction ($t_0$), mass-ratio
($q$), inclination ($i$), the radii scaled by the semi-major axis
($r_{1,2}$), the effective temperature of both components
($T_\mathrm{eff; 1,2}$), and the velocity scale ([$(K_{1}$ + $K_{2}$]
$\sin{i}^{-1}$).

Ellipsoidal variations alone are not sufficient to fully constrain the
binary parameters, and we therefore add priors and other model
constraints.  We use Gaussian priors on the sdB temperature as derived
from the spectroscopy, surface gravity, and radial velocity
semi-amplitude. We fix the limb and gravity darkening parameters using
the tabulated values by \citet{Gianninas2013} and fix the beaming
exponent to 1.62 \citep{LoebGaudi2003,Bloemen2011}.  For the white
dwarf we use the approximation of the mass-radius relation of Eggleton
from \citet{VerbuntRappaport1988} to constrain the radius. First, the
white dwarf radius cannot be smaller than the mass-radius
relation. Second, we use a Gaussian prior to constrain the relative
size of the white dwarf compared to the zero-temperature white dwarf
relation to 10\%.  To find the best parameter values and
uncertainties, we use a Markov-Chain Monte Carlo (MCMC) method as
implemented in \textit{emcee} \citep{ForemanMackey2013}. Before we
fitted the model, we detrended and rescaled the uncertainties of the
three individual lightcurves. We fit the combined lightcurve
using 512 walkers and 2000 generations and show the resulting best fit
and the folded light curve in Fig. \ref{OWJ0815-lcurve-fit} and the
resulting best fit parameters with their associated errors in Table
\ref{OW0815-fits}.

The mass of the white dwarf is $\sim$0.7\Msun\, ($T_{\rm
  eff}\sim14,000$ K) whilst the mass of the sdB star is
$\sim$0.34\Msun\, ($T_{\rm eff}\sim26,500$K).  The component masses
and orbital period of (73.7 min), make it similar to the well known
white dwarf, sdB binary CD--30$^{\circ}$\,11223 which has an orbital
period of 70.5 min \citep{Vennes2012} {and PTF1 J2238+7430 the
  recently discovered white dwarf, sdB binary \citep{Kupfer2022}. The
recent catalogue of post common enevlope binaries of
\citet{Kruckow2021} lists five white dwarf, sdB binaries with an
orbital period shorter than 80 min: OW J0815-3421 is now the sixth
such binary. The mass of the component stars are similar to ZTF
J2320+3750 (55.25 min) \citep{Burdge2020a} and ZTF J2055+4651 (56.35
min) \citep{Kupfer2020}.

The finding that OW\,J0815--3421 is a binary indicates that
  medium resolution spectroscopic observations made over several days
  could reveal evidence for binarity in the other sources identified
  in this survey (c.f. the comments of \citet{Byrne2020} on the
  Roche-lobe overflow model for the formation of BLAPs).

\begin{table}
\begin{center}
\begin{tabular}{lr}
\hline
t$_0$ (HJD) & $2459267.432196$(26) \\[1mm]
$P$ (d) & $0.0511843$(16)\\[1mm]
$q$  & $0.493^{+0.061}_{-0.072}$\\[1mm]
$i$  ($^\circ$) & $83.14^{+4.60}_{-10.55}$\\[1mm]
$M_1$ (\Msun) & $0.707^{+0.086}_{-0.082}$\\[1mm]
$M_2$ (\Msun) & $0.343^{+0.076}_{-0.066}$\\[1mm]
$R_1$ (\Rsun) & $0.01189^{+0.0012}_{-0.0011}$\\[1mm]
$R_2$ (\Rsun) & $0.216^{+0.018}_{-0.015}$\\[1mm]
$a$ (\Rsun)  & $0.590^{+0.027}_{-0.027}$\\[1mm]
$T_1$ (K) & $14077^{+11290}_{-6233}$\\[1mm]
$T_2$ (K) & $26461^{+1052}_{-965}$\\[1mm]
$\log g_{2}$ & $5.44^{+0.021}_{-0.028}$\\[1mm]
\hline
\end{tabular}
\end{center}
\caption{The system parameters derived for OWJ0815--3421 using {\tt Lcurve} to 
  model the light curve.}
\label{OW0815-fits}
\end{table}

\begin{figure}
  \centering
  \hspace{1cm} \includegraphics[width=8.5cm]{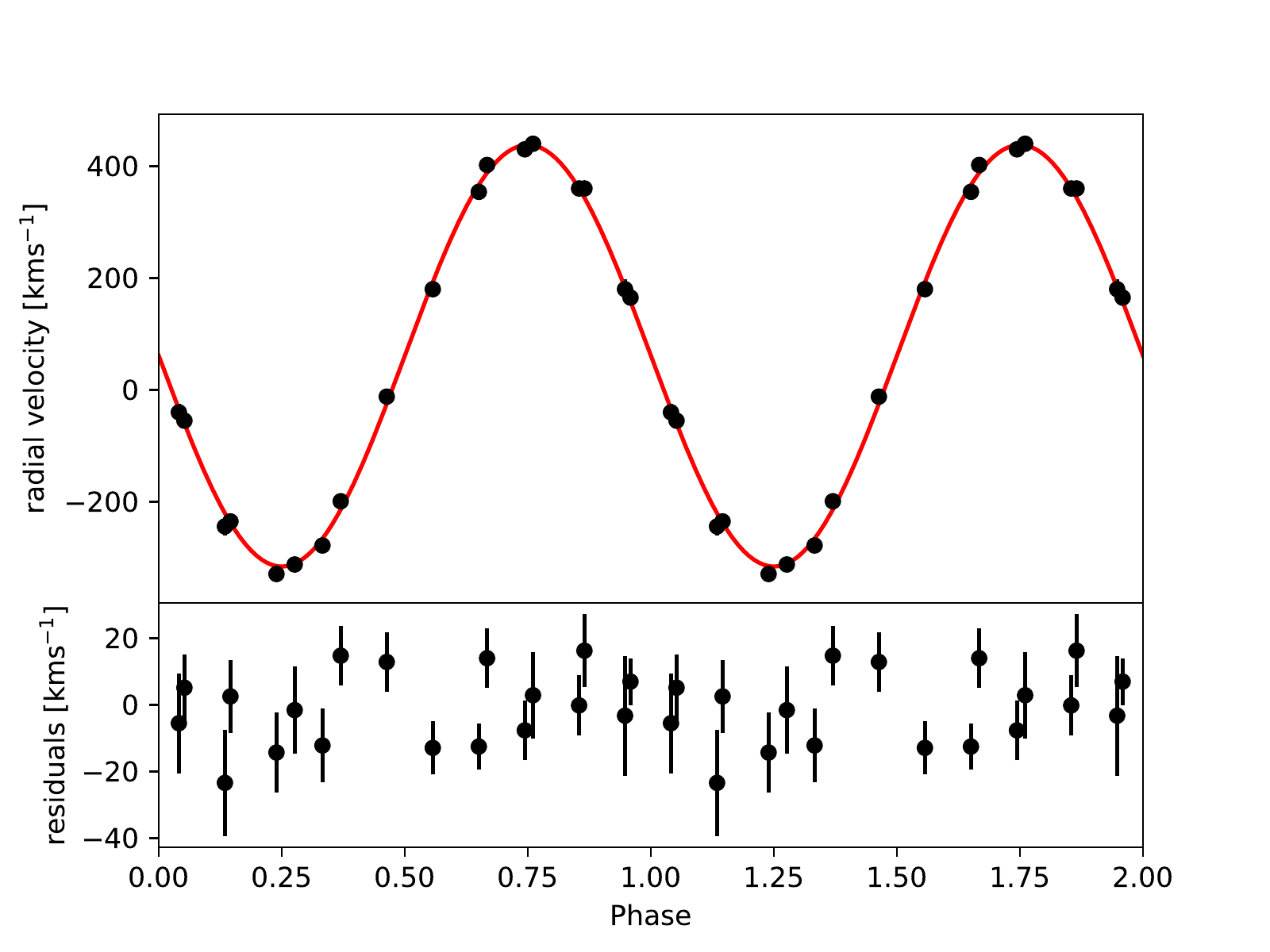}\hfill
  \caption{Upper panel: The radial velocity data of OW\,J0815--3421
    phased on the binary orbital period of 73.7 min together with the
    best fit sinusoid. The lower panel: the residuals to the fit. Two orbits are shown for better visualisation.}
\label{OWJ0815-spec-rv-fit}
\end{figure}

\begin{figure}
  \centering
  \hspace{1cm} \includegraphics[width=8.5cm]{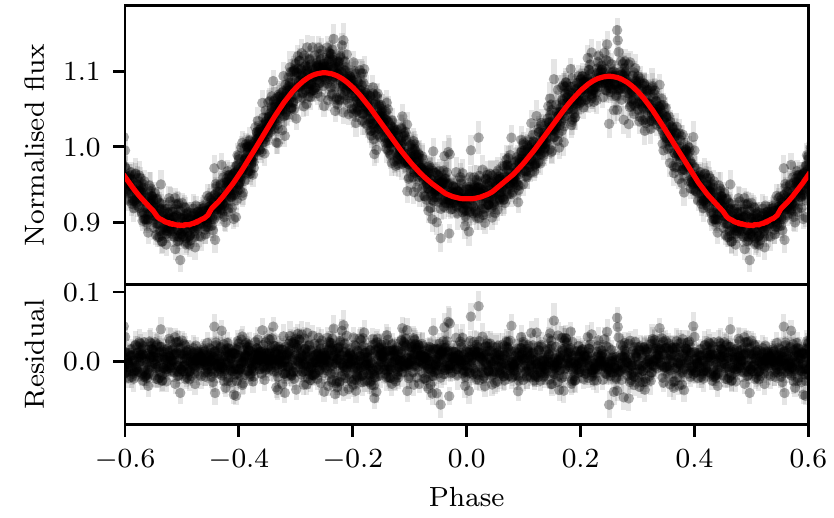}\hfill
  \caption{The data of OW\,J0815--3421, shown in black, has been
    phased on a period of 73.71 min. The red solid line shows the fit
    to the data made using {\tt Lcurve}. The data have been repeated
    to cover two cycles and phase 0.0 represents the secondary
    minimum.}
\label{OWJ0815-lcurve-fit}
\end{figure}

\begin{table*}
\begin{center}
\begin{tabular}{clcccll}
  \hline
  ID & Name & $T_{\rm eff}$ (K) & $\log g$ & $\log n_{\rm (He/H)}$ & Classification & New designation \\
  \hline
 1 & OW J0820-3301 & $37\,400 \pm 1500$ & $5.71 \pm 0.11$ & $-2.5 \pm 0.3$ & V361\,Hya & \\
 3 & OW J1812-2938 & $30\,600 \pm 2500$ & $4.67 \pm 0.25$ & $-2.1 \pm 0.2$ & BLAP  & OW-BLAP-1 \\
 4 & OW J1819-2729 & $12\,300 \pm \ \ 500$  & $4.81 \pm 0.20$ & $-0.6 \pm 0.2$ & sdAV  & \\
 5 & OW J1811-2730 & $27\,300 \pm 1500$ & $4.83 \pm 0.20$ & $-0.7 \pm 0.1$ & BLAP & OW-BLAP-2 \\
 6 & OW J1810-2516 & $29\,900 \pm 3500$ & $4.16 \pm 0.40$ & $-0.8 \pm 0.3$ &  BLAP & OW-BLAP-3 \\
 7 & OW J1758-2716 & $27\,300 \pm 2000$ & $4.20 \pm 0.20$ & $-0.8 \pm 0.2$ & BLAP & OW-BLAP-4\\
 9 & OW J0815-3421 & $26\,500 \pm 1000$ & $5.68 \pm 0.15$ & $-2.8 \pm 0.2$ & sdB & \\
\hline
\end{tabular}
\end{center}
\caption{The results of spectral fits to optical spectra of the
  candidate BLAPs, including the best fit effective temperature,
  surface gravity and helium-to-hydrogen abundance ratio.  The
  penultimate column shows the likely variable star classification.}
\label{spec-fits}
\end{table*}

\section{Discussion}
\label{discussion}

Of the nine sources which we identified as candidate BLAPs (Table
\ref{variables}), four have characteristics consistent with being a
BLAP. The key was obtaining spectra with sufficient spectral
resolution ($R\sim$2000) to be able to fit the absorption lines and
hence obtain sufficiently constrained values for temperature and
gravity. Follow-up high time-resolution photometry was essential in
identifying OW\,J0815--3412 as a very short period binary containing a
sdB star and also identifying the notch feature in four objects, two
of which are BLAPs. We note that the derived periods from follow-up
photometry are consistent with the initial periods derived from the OW
survey. We now discuss all sources reported in this study in the wider
context.

\subsection{Comparison of BLAPs with other short period variables using the Gaia HRD}

The simplest and most readily available means of comparing the colour
and luminosity of different types of sources is through the Gaia HRD
based on ($(BP-RP), MG$) data. Since the OW and the ZTF surveys
observed fields close to the Galactic plane, the extinction can be
high and can show changes on relatively small spatial scales
($<1^{\prime}$). We therefore de-redden the Gaia ($(BP-RP), MG$) data
of variable stars of different types using {\tt Bayestar19}
\citep{Green2019}.

There are a number of different classes of variable stars which lie in
the region of the Gaia HRD where BLAPS have been found
\citep{Ramsay2018,McWhirterLam2022}. For instance, strongly magnetic
Cataclysmic Variables (also known as Polars) can show high (up to 1
mag) variation over their orbital period, but their orbital periods
are \gtae80 min. In contrast, the interacting double degenerate AM CVn
binaries have orbital periods \ltae1 hr but have amplitudes $<$0.1
mag. We therefore include for comparison a number of short period
binaries (6.9< $P_{\rm orb}<$56.4 min) which consist of degenerate
stars and can show high amplitudes of variability, many of which also
show eclipses \citep{Burdge2020a,Burdge2020b} and a small sample of
high amplitude sdB stars\footnote{PG\,1605+072, Balloon\,90100001,
PG\,1716+426, BPS\,BS\,16559--0077, PB\,7032} which have high
amplitudes and short periods. In addition we add the high gravity
  BLAPs reported by \citet{Kupfer2019} and also the high gravity `sdB'
  pulsators reported in \citet{Kupfer2020}. Since the periods of these
  systems are similar, with the latter sample showing only slightly
  lower amplitudes, we class both these samples as high gravity BLAPs,
  although they may have a different internal structure.  We also
take the sources originally reported by \citet{Pietrukowicz2017} and
have Dec$>-30^{\circ}$ and hence the reddening can be determined using
{\tt Bayestar19}. In Fig. \ref{variables-gaia-hrd} we show the
location of these stars in the ($(BP-RP)_{o}, M_{G_o}$) plane and for
comparison stars which are within 50 pc of the Sun, which we assume
have negligible reddening.

As can be seen, the sources cover a wide range in the Gaia HRD,
encompassing the area bluer than the main sequence but more luminous
than the main white dwarf cooling tracks. The BLAPs have a wide range
of $M_{G_o}$ although we caution that because the error on their
parallax is generally high the resulting uncertainty on $M_{G_o}$ can
be up to $\sim$1 mag (c.f. Table \ref{variables}). The compact
binaries tend to be less luminous compared to the other classes with
those being more luminous having an sdB star as a binary
component. Stellar pulsators appear to be more luminous. However, it
is not possible to determine the nature of any given source based on
its location purely in the Gaia HRD.

Fortunately, for seven of the sources in this study we have been able
to determine their temperature and gravity (Table \ref{spec-fits}). In
Fig. \ref{Teffvlogg} we show the location of the OW BLAPs, the BLAP
sample from \citet{Pietrukowicz2017} and the high gravity BLAPs from
\citet{Kupfer2019}. We also show the evolutionary tracks for stars of
different mass taken from \citet{Kupfer2019}.  The OW BLAPS share a
similar part of the \teff,\logg\, plane (relatively low gravity)
whilst the high gravity BLAPS have a shorter pulsation period and high
gravity.

\subsection{Notch feature at maximum brightness}

\citet{Macfarlane2017a} identified a dip in the light curve of
OW\,J1758--2716 occuring near maximum light which was compared to
features seen in `Bump' Cepheid variables. These bumps are features in
the phase folded light curve which occur on the rise to maximum
brightness or in the descent from brightness. The phase of where this
occurs is known as the Hertzsprung Progression
\citep{Hertzsprung1926,Bono2000} and depends on stellar mass and
therefore the observed period. There are Cepheids which show a dip at
flux maximum very similar to what we observe in OW\,J1758--2716, see
for example RU Dor (Fig 9, \citet{Plachy2021}). It is likely that Bump
features are due to a resonance between the fundamental pulsation mode
and the second overtone (e.g. \citet{GastineDintrans2008} and
references therein).

We do not have additional photometric observations since those
reported by \citet{Macfarlane2017a} which showed photometry covering a
timeline of $\sim$160 min. Further observations would be required to
search for a second over-tone in the light curve. However, as shown in
Fig. \ref{avphot}, OW\,J1811--2730 and OW\,J1826--2812 show hints for
a slight bump in their folded light curves on the rise to maximum
brightness. There is no sign of bump features in the high gravity
BLAPs identified by \citet{Kupfer2019}.

\subsection{sdA pulsators}

High-gravity A stars (sdA's) have been widely discussed since being
identified in the Sloan Digital Sky Survey with H-dominated spectra, $
9000 > T_{\rm eff} > 6500$ K), and $ 6.5 > \log g > 5.5 $
\citep{Kepler2016}.  Whilst purely a spectroscopic classification,
they have been identified with binaries of a subdwarf and a main
sequence object of type FGK; main sequence A stars with a high surface
gravity; or extremely low-mass white dwarfs (ELMs) or their precursors
(pre-ELMs). Of these only the latter (ELMs and pre-ELMs) stand
scrutiny \citep{Pelisoli2016,Bell2018}.

\citet{Bell2018} in their search for pulsating sdA stars found
  three pre-ELM candidates which have temperatures ($\sim$8,000 K)
  which are slightly cooler than that measured for
  OW\,J1819--2729. They indicate that pre-ELM stars might undergo CNO
  flashes on a short timescale.  \citet{Byrne2020} studied the
pulsation properties of pre-ELMs and ELMs as they evolve from the
subgiant branch, where they must lose mass to a close companion, onto
the ELM white dwarf sequence.  Their models included chemical
diffusion, and demonstrated how opacity from concentrated iron-group
elements can switch on pulsations in pre-ELM models exactly where
BLAPs had previously been identified
\citep{Pietrukowicz2017,Kupfer2019}.  A few models with core masses
$0.255 \leq M_{\rm core} \leq 0.305$ \Msun\ show a late hydrogen-shell
flash after becoming an ELM.  This flash inflates the white dwarf
causing a loop in the evolution track which passes through the $T_{\rm
  eff}$,\logg\, plane in the region occupied by OW\,J1819--2729.
During this time the models are also unstable to pulsation, with
periods approximately between 200 and 4000\,s ($\log g$ between 6.0
and 4.0).

Difficulties with this argument include the short time that models
spend in such flash-driven loops, and the expectation that, in order
to become an ELM, the agency of a companion is required. Nevertheless,
further evidence to test the notion that OW\,J1819--2729 is a
shell-flashing ELM should be sought.

\subsection{Space Density of BLAPs}

The OW survey covered 404 distinct fields each covering one square
degree.  We have identified four stars which have short photometric
periods and high amplitudes with followup spectroscopic observations
indicate they are BLAPs, implying one BLAP per $\sim$0.01
deg$^{-1}$. \citet{Byrne2021} made predictions of how many BLAPs would
be expected in our Galaxy using the Binary Population Spectral
Synthesis code \citep{Eldridge2017}. They found the numbers were very
dependent on Galactic latitude (concentrated towards the plane) and
longitude (highest the Bulge) although regions of high extinction
could make this very variable. Regions with the highest concentrations
of predicted BLAPs reach $\sim$0.3 deg$^{-1}$ at a depth of $g$=20
mag, but a median of 0.001 deg$^{-1}$ at $b$=$0^{\circ}$. Moving to
$b$=$15^{\circ}$ results in a very low number of BLAPs being
predicted. Our search for BLAPs in the OW survey did not reach as
faint as $g$=20 mag but with an observed number density of $\sim$0.01
deg$^{-1}$ down to $g\sim19$, our findings appear reasonably
consistent with the predictions of \citet{Byrne2021}, giving good
grounds to suggest that 12,000 BLAPs may exist in our Galaxy.

\begin{figure}
  \vspace{3mm}
  \includegraphics[width=7.5cm]{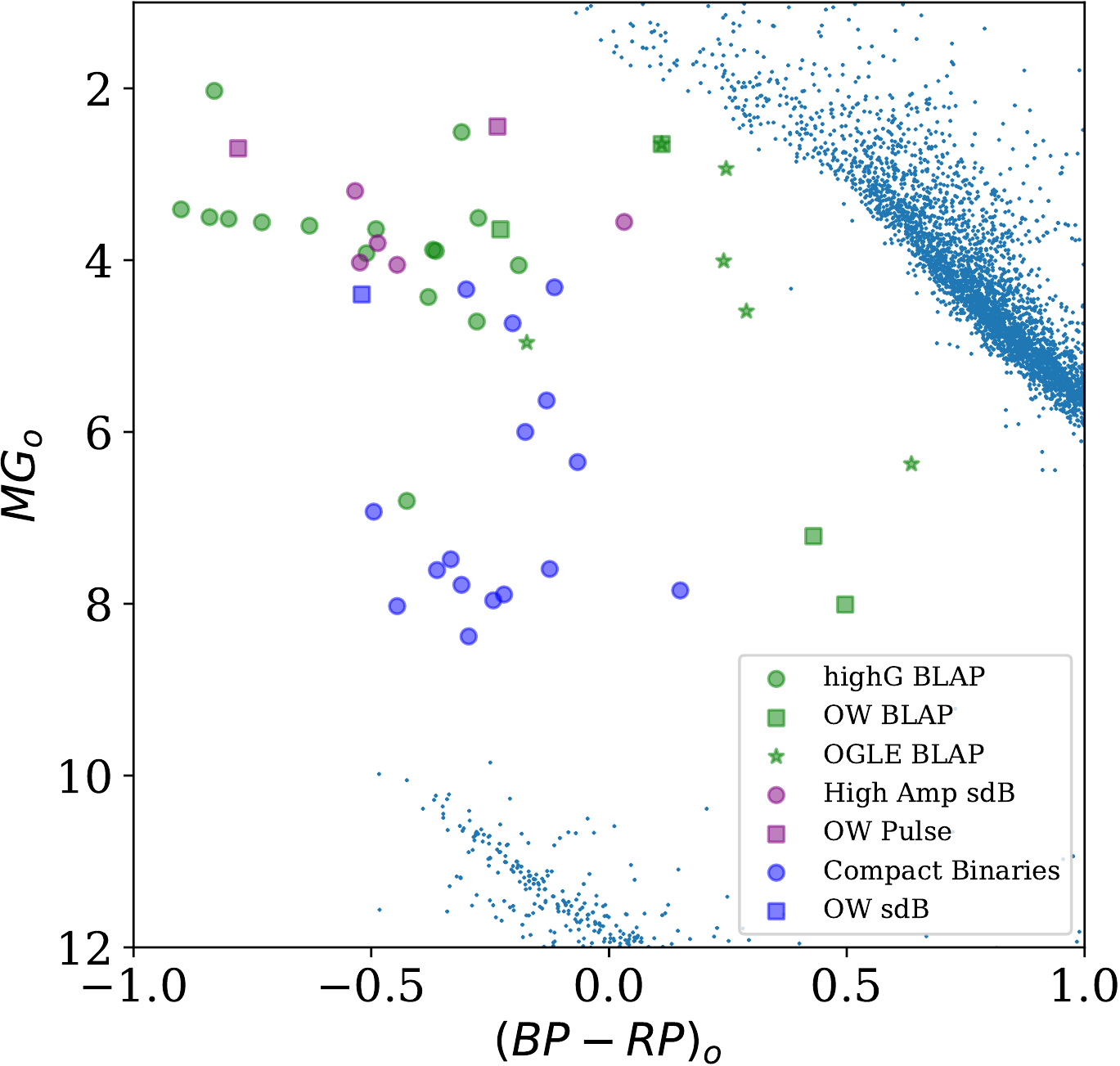}
  \caption{The observed $(BP-RP)_o$, $MG_{o}$ colour-absolute
    magnitudes for the BLAPs and other high amplitude pulsators as
    identified in the OW survey. For comparison we also show the
    position of the high gravity BLAPs reported in \citet{Kupfer2019,Kupfer2021};
    compact binaries identified in \citet{Burdge2020a,Burdge2020b};
    and high amplitude sdB stars.}
\label{variables-gaia-hrd}
\end{figure}

\begin{figure}
\includegraphics[width=8.4cm]{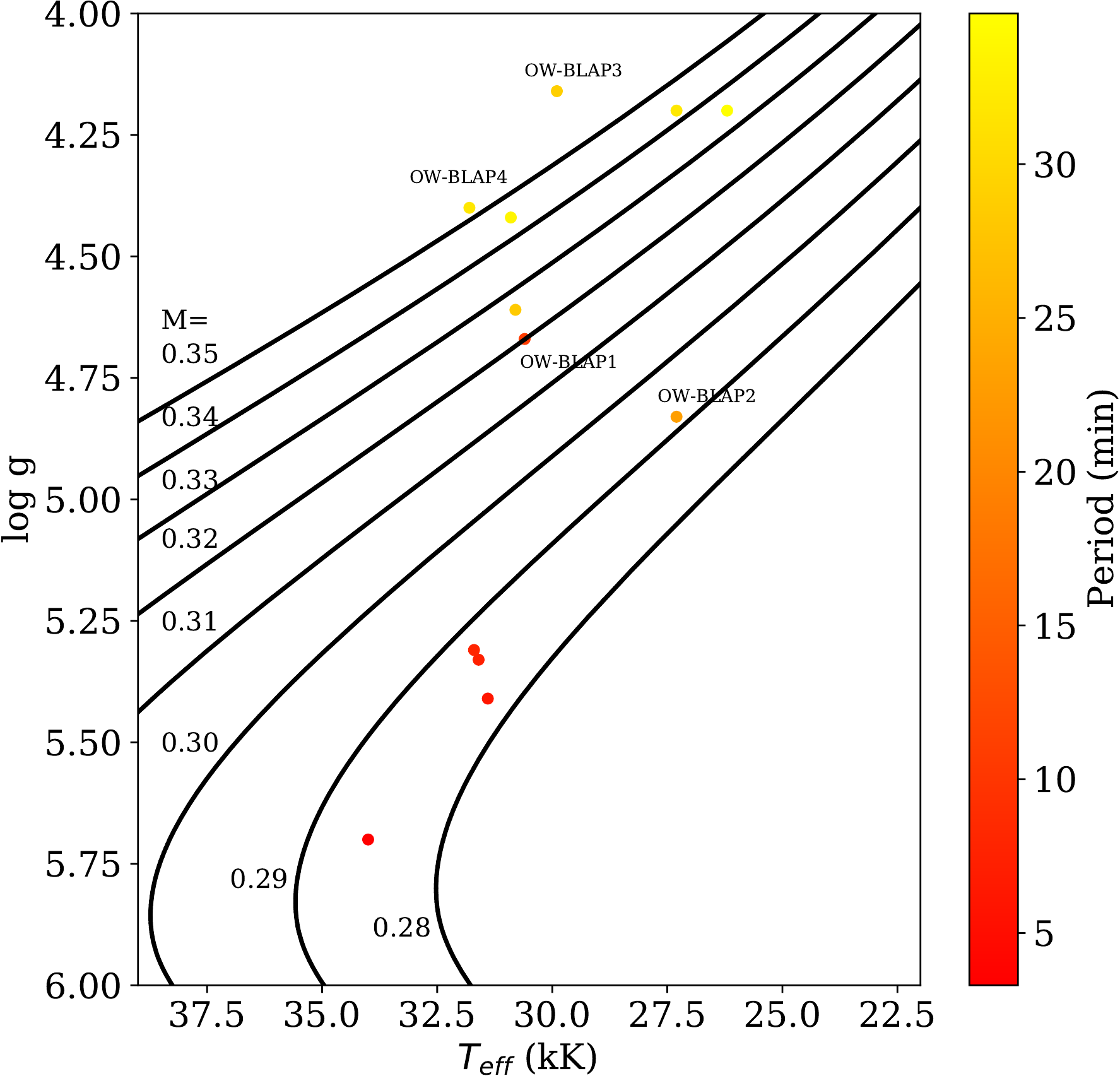}\hfill
  \caption{The solid lines are predicted evolutionary tracks for BLAPs
    in the T$_{eff}$, log $g$ plane and taken from \citet{Kupfer2019}
    for stars of different masses. The BLAPs identified in the OW
    survey are indicated where the colour of the symbol indicates the
    stars pulsation period. The shorter period BLAPs come from the
    sample of \citet{Pietrukowicz2017} while the longer period (and
    high gravity) systems are from the sample of
    \citet{Kupfer2019}. The OW BLAPs share a similar part of the plane
    as the lower gravity sample of \citet{Pietrukowicz2017}.}
\label{Teffvlogg}
\end{figure}

\section{Conclusion}

We have used data taken as part of the OmegaWhite survey to search for
short period, high amplitude variables which may be BLAPs: we
identified nine stars which we initially classed as candidate
BLAPs. Using a combination of followup photometry, spectroscopy and
Gaia data we were able to identify four of these variables as BLAPs,
one of these being previously identified as a BLAP by
\citet{Pietrukowicz2017}. Overall they have similar properties to the
BLAPs identified by \citet{Pietrukowicz2017} rather than the high
gravity BLAPs identified by \citet{Kupfer2019,Kupfer2021}.

We have also identified OW\,J0815--3421 as a binary system containing
an sdB and a low mass star. With an orbital period of 73.7 min, it is
one of half a dozen white dwarf - sdB star binaries with orbital
periods shorter than 80 min. A more detailed investigation is
  required to determine if it is a double-detonation thermonuclear
  supernova progenitor like PTF1 J2238+7430 \citep{Kupfer2022}.
OW\,J1819--2729 is a pulsating sdAV star which has interesting
properties which makes future modelling an interesting future
study. One further source, OW\,J0820--3301, is a V361\,Hya pulsator
which has an extended nebula which is spatially nearby. Although at
this point we cannot be certain it is physically associated with the
short period variable, further observations using an integral field
spectrograph maybe able to determine if there is any association
between the nebula and the variable star.

Finally we note that high cadence observations of these short period
variable stars have revealed detailed structure in their folded light
curve. We make comparison with other types of pulsating variable stars
which show similar features in their light curves. We encourage
stellar modellers to reproduce such features using appropriate
models.

\section{Acknowledgments}

The targets identified in this paper were identified in data obtained
using the ESO VST and OmegaCam under proposals: 088.D-4010;
090.D-0703; 091.D-0716; 092.D-0853; 093.D-0753; 093.D-0937;
094.D-0502; 095.D-0315; 096.D-0169; 097.D-0105; 098.D-0130; 099.D-0164
and 0100.D-0066. SALT spectroscopic data were obtained under
proposals: 2015-2-SCI-035; 2016-1-DDT-004; 2016-1-MLT-010;
2016-1-SCI-015 and 2017-2-SCI-051.  This paper uses observations made
at the South African Astronomical Observatory and the South African
Large Telescope and we thank the staff at both for their expertise in
obtaining these data.  Based on observations made with the WHT
(programme ID: W/16B/N4) operated on the island of La Palma by the
Isaac Newton Group of Telescopes in the Spanish Observatorio del Roque
de los Muchachos of the Instituto de Astrof\'{i}sica de Canarias. PAW
acknowledges support from UCT and the NRF. TK acknowledges support
from the National Science Foundation through grant AST \#2107982, from
NASA through grant 80NSSC22K0338 and from STScI through grant
HST-GO-16659.002-A.  PJG acknowledges support from NOVA for the
original OmegaWhite observations in the Dutch VST/Omegacam GTO
observations. PJG is supported by NRF SARChI Grant 111693. Armagh
Observatory and Planetarium is core funded by the Northern Ireland
Executive through the Department for Communities. UH and AI
acknowledge funding by the Deutsche Forschungsgemeinschaft (DFG)
through grants HE1356/70-1 and HE1356/71-1. We thank Kinwah Wu for a
useful discussion on probabilities of stars being spatially nearby and
Evan Bauer who generated the evolutionary tracks shown in
Fig. \ref{Teffvlogg}.  We thank Steven H\"ammerich for providing us
with synthetic spectra for solar metal composition. We thank the
referee for a useful report.

\section*{Data Availability}

The images obtained as part of the OW survey can be accessed through
the ESO portal. Light curves and spectra can be obtained via a
reasonable request from the authors.

\end{document}